\documentclass[12pt]{article}
\usepackage{amsmath}
\usepackage{graphicx}
\usepackage{natbib}
\usepackage{url} % not crucial - just used below for the URL 
\usepackage[english]{babel}
\usepackage{bbm, dsfont}
\usepackage{array,colortbl,xcolor}
\usepackage{lipsum}
\usepackage{tabu}
\usepackage{lscape}
\usepackage{afterpage}
\usepackage{lscape}
\usepackage{placeins}
\usepackage{mathtools}
\usepackage{lineno}
\usepackage{epstopdf}
\usepackage{setspace}
\usepackage{pbox}
\usepackage{booktabs}
\usepackage{rotating}
\usepackage{tabularx}
\usepackage{caption}
\usepackage{subcaption}
\usepackage{array}
\usepackage{epstopdf}
\usepackage{prettyref}
\usepackage{multirow}
\usepackage{lipsum}
\usepackage{endnotes}
\usepackage{adjustbox}
\usepackage{longtable,tabulary}
\usepackage{ltablex}
\usepackage[T1]{fontenc}
\usepackage{inputenc}
\usepackage[toc,page]{appendix}
\usepackage{hyperref}
\hypersetup{
    colorlinks = true,
    linkcolor = blue,
    anchorcolor = blue,
    citecolor = blue,
    filecolor = blue,
    pagecolor = blue,
    urlcolor = blue
}
\usepackage{pifont}
\usepackage{fleqn}
\usepackage{txfonts}
\usepackage{longtable}
\usepackage{makecell}
\usepackage{rotating}
\usepackage{color}
\usepackage{amsfonts,epsfig,epstopdf,array}
\usepackage{booktabs,dcolumn}
\usepackage[ruled]{algorithm2e}

\usepackage{nomencl}
\makenomenclature
\DeclareMathOperator*{\argmin}{arg\,min}
\DeclareMathOperator*{\argmax}{arg\,max}

 \relax

\newcommand{\blind}{0}

\graphicspath{{Figures/}{TexFiles/}}

\newcommand{\bftau}{\boldsymbol{\tau} }

\newcommand{\bfX}{\boldsymbol{X}}
\newcommand{\bfx}{\mathbf{x}}
\newcommand{\bfmu}{\boldsymbol{\mu} }
\newcommand{\bfbeta}{\boldsymbol{\beta} }

\newcommand{\bfSigma}{\boldsymbol{\Sigma} }
\newcommand{\bftheta}{\boldsymbol{\theta} }
\newcommand{\bfPsi}{\boldsymbol{\Psi} }
\newcommand{\bfTheta}{\boldsymbol{\Theta} }

\newcommand{\bfH}{\boldsymbol{H}}

\newcommand{\bfS}{\boldsymbol{S}}
\newcommand{\bfW}{\boldsymbol{W}}

\newcommand{\bflambda}{\boldsymbol{\lambda}}
\newcommand{\mS}{\mathbb{S}}

\DeclareMathOperator{\tr}{tr}
\DeclareMathOperator{\diag}{diag}

\usepackage{ntheorem}
\newtheorem{thm}{Theorem}
\newtheorem{corollary}[thm]{Corollary}
\newtheorem{theorem}[thm]{Theorem}
\newtheorem{lemma}[thm]{Lemma}
\theoremstyle{nonumberplain}
\newtheorem{proof}{Proof}

\begin{document}

\def\spacingset#1{\renewcommand{\baselinestretch}%
{#1}\small\normalsize} \spacingset{1}

%%%%%%%%%%%%%%%%%%%%%%%%%%%%%%%%%%%%%%%%%%%%%%%%%%%%%%%%%%%%%%%%%%%%%%%%%%%%%%

\if0\blind
{\title{\bf Sparse Graphical Modelling via the sorted $\ell1$ - Norm\thanks{The opinions expressed in this article are those of the authors and do not necessarily reflect the views of La Francaise Systematic Asset Management or any of its affiliates.}}
  
  \author{Riccardo Riccobello\thanks{MB acknowledges the  support of the Swedish Research Council, grant no.
2020-05081.}  \hspace{.2cm}\\
    Department of Economics and Management, University of Trento,\\
    Ma{\l}gorzata Bogdan \\
    Department of Mathematics, University of Wroc{\l}aw\\
    Department of Statistics, Lund University,\\
    Giovanni Bonaccolto\\
    Faculty of Economics and Law, Kore University of Enna,\\
    Philipp J. Kremer\\
    Business School, EBS University for Business and Law,\\
    Sandra Paterlini\\
    Department of Economics and Management, University of Trento\\
    and\\
    Piotr Sobczyk\\
    Department of Mathematics, Wroclaw University of Science and Technology}
    
  \maketitle
} \fi

\if1\blind
{
  \bigskip
  \bigskip
  \bigskip
  \begin{center}
    {\LARGE\bf Title}
\end{center}
  \medskip
} \fi

\bigskip
\begin{abstract}

Sparse graphical modelling has attained widespread attention across various academic fields.  We propose two new graphical model approaches, Gslope and Tslope, which provide sparse estimates of the precision matrix  by penalizing its sorted $\ell_1$-norm, and relying on Gaussian and T-student data, respectively. We provide the selections of the tuning parameters which provably control the probability of including false edges between the disjoint graph components and empirically control the False Discovery Rate for the block diagonal covariance matrices.
In extensive simulation and real world analysis, the new methods are compared to other state-of-the-art sparse graphical modelling approaches. The results establish Gslope and Tslope as two new effective tools for sparse network estimation, when dealing with both Gaussian, t-student and mixture data.

\end{abstract}

\noindent%
\vfill

\newpage
\spacingset{1.5} % DON'T change the spacing!

\section{Introduction}\label{sec:introduction}
Massive data sets are nowadays routinely collected in many fields of science and business.
Acquiring the knowledge from such huge data collections usually relies on discovering some hidden patterns,
like the dependency structure between different variables. One set of tools to recover this structure is provided by the probabilistic graphical models,  which use graphs to encode the relationships between different variables (see, e.g. \cite{Lauritzen1996}). 

In Graphical Models the relationship between variables is described by the  graph (or network)  $\mathtt{G}=\left(\mathtt{V},\mathtt{E} \right)$, where the elements of the $\mathtt{V}$ and $\mathtt{E}$ sets are the vertices (or nodes) and edges (or links) of $\mathtt{G}$, respectively. The vertices of $\mathtt{G}$ correspond to variables, or, in other words, entries of the $p \times 1$ random vector $\bfX =\left[   X_1,\ldots , X_p \right]^\prime$. 
%For the sake of simplicity, we set $\{X_1, \ldots , X_p\} \equiv \{1, \ldots , p\}=\mathtt{V}$. 
%Therefore, $\mathtt{G}=\left(\mathtt{V},\mathtt{E} \right)$ provides a graphical tool to study the joint distribution of these returns, 
In this article we focus on undirected graphs, also known as Markov random fields or Markov networks,  
%so that $(ij)=(ji) \in \mathtt{E}$, where $i,j \in \mathtt{V}$ and $i \neq j$.  In other words, the edges in $\mathtt{E}$ reflect mutual relationships between two different vertices $i$ and $j$. 
%In undirected graphs,  
where the absence of an edge between two vertices $i$ and $j$ means that $X_i$ and $X_j$ are conditionally independent, given the other variables of the random vector $\bfX$; that is,  $X_i \perp X_j |  \bfX_{-(i,j)}$, where $\bfX_{-(i,j)}$ denotes the $\bfX$ vector without $X_i$ and $X_j$
%, for $1 \leq i,j \leq p$ and $i \neq j$
\citep{Dempster1972, Murphy2012, Hastie2017}. 
%We display a simple example of undirected graph in Figure \ref{fig:graphexample}, where $\mathtt{V}=\{1,2,3,4,5,6\}$ and $\mathtt{E}=\{12,23,31,45\}$. We can see, for instance, that vertices 5 and 6 are not connected by an edge in Figure \ref{fig:graphexample}, so that $X_5 \perp X_6 | \textbf{X}_{-(5,6)}$. 
 
%\begin{figure}[htb!]
%\hbox{\hspace{4cm}\includegraphics[scale=0.8]{graph.pdf}}
%\caption{\footnotesize{Simple example of undirected graph.}}
%\label{fig:graphexample}
%end{figure} 
 
If the structure of the undirected graph is not known \textit{a priori}, we need to perform graphical model selection; that is, to %estimate
identify the edges belonging to $\mathtt{E}$. Specific assumptions about the distribution of $\bfX$ are then required. For example, in Gaussian Graphical Models (GGM),  it is assumed that $\bfX$ follows a 
%zero-mean 
multivariate normal distribution  $\mathcal{N}_p\left(\bfmu, \bfSigma \right)$, 
where $\bfmu=\left[\mathbb{E}[X_1]\;\cdots \; \mathbb{E}[X_p]\right]^\prime$ and 
$\bfSigma=\mathbb{E}\left[\bf(X-\mu)  \bf(X-\mu)^\prime \right]$ is the covariance matrix of the random variables $X_1,\ldots , X_p$.
%Without loss of generality, we can assume that $\bfmu=\textbf{0}$, where $\textbf{0}$ denotes a zero vector \citep[see, among others,][]{JonesWest2005, Malioutov2006, Peterson2015, LeeSobczykBogdan2019}. 

% due to its appealing theoretical properties. 
% We then estimate Gaussian graphical models on $\bfX \sim  assuming that the expected value vector is $\textbf{0}$, 
%without loss of generality \citep{LeeSobczykBogdan2019}. 
%\red{Remove mu and S as eq 1 and 2 and have them as intext equation} 
%Assume that we have observed $n$ realizations of $\bfX$, denoted as $\bfx^{(j)} = \left[x^{(j)}_1 \; x^{(j)}_2 \cdots  x^{(j)}_p \right]$

Gaussian graphical models are also known as covariance selection or concentration graph models \citep{TorriGiacomettiPaterlini2019}, since they rely on the inverse of the covariance matrix $\bfTheta = \bfSigma^{-1}$ (i.e. the precision or concentration matrix) to determine the graph structure. Specifically, $\bfTheta$ provides information about the partial covariance between $X_i$ and $X_j$ conditional on $\bfX_{-(i,j)}$, for $i,j=1,\ldots , p$ and $i \neq j$;  %\red{DEFINE $\bfX_{-(i,j)}$ } \textcolor{green}{giovanni: we already defined it above, at the end of the first paragraph.} 
 $X_i$ and $X_j$ are conditionally independent, given the variables in $\bfX_{-(i,j)}$, if and only if $\bfTheta_{i,j}=0$ \citep{Lauritzen1996,Hastie2017}. 
 %More generally, the partial correlation of $X_i$ and $X_j$ can be defined as the following function of $\theta_{i,j}$: $\eta_{i,j} = - \frac{\theta_{i,j}}{\sqrt{\theta_{i,i} \theta_{j,j}}}$, or in matrix notation: $\bfR=-\bfUpsilon \bfTheta \bfUpsilon$, where $\bfR$ is the partial correlation matrix and $\bfUpsilon=\diag \left[(\theta_{1,1})^{-1/2},\ldots, (\theta_{p,p})^{-1/2} \right]$; see, among others, \cite{Lauritzen1996}. 
 Moreover, the Gaussian graphical model can be represented as the set of $p$ multiple regression models, where for each $i\in\{1,\ldots,p\}$
\begin{equation}\label{eq:localreg}
 X_i = \sum_{j\neq i} X_{j} \beta_{i,j} + \nu_i, \;\; \nu_i \sim \mathcal N(0, \sigma_i^2)\;\;.
\end{equation}
In this representation $X_i$ and $X_j$ are conditionally independent if and only if $\beta_{i,j}=0$ (see e.g., \cite{Anderson2003}).
Due to this conceptual clarity Gaussian Graphical Models are probably the most popular group of undirected graphical models and are nowadays routinely used for structure-discovery in many fields, like neuroimaging, genetics, or finance (see e.g., \cite{FriedmanHastieTIbshirani2008, Wang2011, Ryali2012, Mohan2012, Belilovsky2016, Zhao2019}). 
 
Given $n$ realizations of the random vector $\bfX \sim \mathcal{N}_p \left(\bfmu, \bfSigma \right)$, denoted as $\bfx^{(1)},\ldots , \bfx^{(n)}$,
%the sample covariance matrix $\bfSigma$ is defined as
%the sample mean vector: $\widehat{\pmb{\mu}}=\frac{1}{n}\sum_{i=1}^{n}\textbf{x}_i$ and 
%: $\bfS= \widehat{\bfSigma}=\frac{1}{n}\sum_{j=1}^{n} \bfx^{(j)} \left( \bfx^{(j)}\right)^\prime$.
the log-likelihood of the data is proportional to 
$$\mathcal{L}\left( \bfTheta \right) = \log \det \bfTheta  -\tr \left( \bfTheta \bfS  \right)\;\;,$$
where $\bfS$ 
%=\frac{1}{n}\sum_{j=1}^{n} \left(\bfx^{(j)}-\hat \mu\right) \left( \bfx^{(j)} - hat \mu\right)^\prime$
is the sample covariance matrix defined as 
\begin{equation}\label{eq:samcov}
\bfS=\frac{1}{n}\sum_{j=1}^{n} \left(\bfx^{(j)}-\hat \mu\right) 
\left( \bfx^{(j)} - \hat \mu\right)^\prime\;, 
\end{equation}
with $\hat \mu=\frac{1}{n} \sum_{j=1}^n \bfx^{(j)}$ (see, among others, \cite{Murphy2012} and \cite{Hastie2017}). The maximization of $\mathcal{L}\left( \bfTheta \right)$, with respect to $\bfTheta$, provides $\bfS^{-1}$ as the maximum likelihood estimator of $\bfTheta$. Nevertheless, this estimator is typically unsatisfactory or ill-defined when $p$ approaches to or is greater than $n$ \citep{PircalabeluClaeskens2020}. In particular, when $p > n$, the covariance matrix $\bfS$ is singular and  $\bfS^{-1}$ does not exist. 

The classical solution to the above issues relies on the application of the penalized likelihood estimators obtained by solving the following optimization problem:
\begin{equation}\label{eq:optimgen}
\widehat{\bfTheta} = \operatorname*{arg\; max}_{\bfTheta> \textbf{0}}\; \mathcal{L}\left( \bfTheta \right) - Pen(\bfTheta)\;\;,
\end{equation}  
where the constraint $\bfTheta > \textbf{0}$ guarantees that the solution is positive-definite and the penalty term $Pen(\bfTheta)$ penalizes the model complexity. The classical model selection criteria, like the Akaike Information Criterion \cite{Aka} or the Bayesian Information Criterion \cite{Schw} directly penalize the number of graph edges. In these cases $Pen(\bfTheta)=f(||\bfTheta||_0)$, where $||\bfTheta||_0$ is the number of nonzero elements of $||\bfTheta||$ and $f(\cdot)$ is an increasing function. As discussed e.g. in  \cite{Rina},  when $p$ is comparable or larger than $n$ then AIC and BIC lead to many false discoveries and need to be replaced by the criteria which penalize the dimension of $\bfX$, like the modified Bayesian Information Criterion (\cite{mBIC, mBIC2}) or the Extended Bayesian Information Criterion (\cite{CC08}). However, similarly as in the multiple linear regression, identifying the model which yields the maximum value of a given model selection criterion is NP-hard. Therefore the $L_0$ penalty is often replaced by some convex penalties like the $\ell_2$ or the $\ell_1$ norms of the vectorized version of $\bfTheta$.

The $\ell_1$-norm penalization is a popular technique for obtaining  sparse estimators in a wide range of statistical problems. It was originally employed by \cite{Santosa86} in geo-physics and by \cite{chen1994basis} in the context of signal-processing. In \cite{Tibshirani1996} $\ell_1$-norm penalty was introduced into the general statistics as the well-known Least Absolute Shrinkage and Selection Operator (LASSO) for selection of important variables in regression models. 
The first application of LASSO to the sparse inverse covariance estimation was proposed by \cite{Meinshausen2006} as the neighborhood selection method. In this approach LASSO is applied separately to solve each of the multiple regression problems (\ref{eq:localreg}). Subsequently, in \cite{FriedmanHastieTIbshirani2008} the graphical lasso (Glasso) was proposed, where $\ell_1$ penalty is used directly with the multivariate normal likelihood. 

Specifically, the Glasso estimation builds upon the following optimization problem:   
\begin{equation}\label{eq:optimGlasso}
\widehat{\bfTheta} = \operatorname*{arg\; max}_{\bfTheta > \textbf{0}}\; \left\lbrace  \log \det \bfTheta  -\tr \left( \bfTheta \bfS  \right) -\lambda \left\Vert \bfTheta \right\Vert_1 \right\rbrace,
\end{equation}  
 where 
%$\left| \textbf{Z} \right|$ and $\tr \left[ \textbf{Z} \right]$ denote, respectively, the determinant and the trace of a given square matrix $\textbf{Z}$; 
%$\left( \log \left| \pmb{\Omega} \right| -\tr \left[\pmb{\Omega} \textbf{S}  \right] \right)$ is the log-likelihood function computed from the multivariate normal distribution, partially maximized with respect to $\pmb{\mu}$ up to constant terms \citep{Murphy2012, Hastie2017}; 
$\left\Vert \bfTheta \right\Vert_1 = \sum_{i \neq j} \left| \theta_{i,j} \right|$ is the $\ell_1$-norm of $\bfTheta$ (i.e. the sum of the absolute values of the entries of $\bfTheta$), whereas $\lambda > 0$ is the tuning parameter which determines the intensity of the penalization. Glasso has been proved to be consistent under certain assumptions \citep{Ravikumar2008} and has received considerable attention in the literature \citep[see, among others,][]{Meinshausen2006, MazumderHastie2012, Murphy2012, Pourahmadi2013, Sojoudi2016, Fattahi2019, TorriGiacomettiPaterlini2019, PircalabeluClaeskens2020}. 

Furthermore, in \cite{Finegold2011} the scale-mixture representation of the t-distribution was used to extend Glasso to handle the heavy-tailed distributions. The new method, so called Tlasso, has proven to be an effective tool for a robust graphical inference in presence of outliers or contaminated data \citep{Finegold2014, TORRI201851, Cribben2019, TorriGiacomettiPaterlini2019}.

Despite its appealing properties, LASSO suffers from relevant shortcomings. For instance, it typically provides biased estimates, overshrinking the retained variables \citep{Fan2001}. Moreover, in the context of multiple regression, it performs a random selection among two or more variables when they are highly correlated \citep{Bondell2008}, which may lead to overlooking some of the important predictors. In the context of the neighborhood selection this may lead to overlooking some of the important graph edges. To solve these problems, several generalizations of LASSO were developed. One of them is the Elastic Net \citep{Zou2005}, which relies on the linear combination of $\ell_1$ and $\ell_2$ penalties and encourages including the groups of correlated predictors. Another extension of LASSO is SLOPE \citep{Bogdan2015}, with the penalty defined by the sorted $\ell_1$ norm (SL1);

\begin{equation}\label{eq: 5.29}
J_{\bflambda}(\bfbeta)=\sum_{i=1}^{p}\lambda_{i} \left|\beta \right|_{(i)},
\end{equation}
where $\bfbeta$ is the vector of the model parameters, whose absolute values are sorted in descending order: $\left|\beta\right|_{(1)} \ldots \geq \left| \beta\right|_{(p)}$, whereas $\bflambda = \left[ \lambda_1 \cdots \lambda_p \right]$ is the sequence of the corresponding tuning parameters, which satisfy the condition $\lambda_{1}\geq...\geq\lambda_{p}\geq 0$.
  
  SLOPE penalty is based on a decaying sequence of tuning parameters, which allows for assigning exactly the same estimated regression coefficients to the groups of variables with a similar influence on the loss function \citep{Figueiredo2014, Schneider2020, Kremer2021} and, in this way, it encourages including the groups of correlated predictors.
Moreover, when predictors are independent, one can select the sequence of tuning parameters so that SLOPE controls the False Discovery Rate (FDR) among the selected regressors \citep{Bogdan2013, Bogdan2015, Virouleau2017, Brzyski2019, Kos2020}. In general, SLOPE exhibits two levels of shrinkage of regression coefficients: i) shrinking towards zero; and ii) shrinking the similar estimates towards each other. This, together with FDR control, allows SLOPE to adapt to unknown signal sparsity and obtain sharp minimax estimation and prediction rates for the orthogonal and the independent gaussian designs \citep{SuCandes2016}. This is in contrast with LASSO, which can obtain sharp minimaxity only by adjusting the tuning parameter $\lambda$ to the unknown sparsity. Consecutively, in a series of works \citep{Bellec2016, Virouleau2017, Bellec2018, Abramovich2019} it was proved that SLOPE attains the minimax estimation rates for the general class of design matrices satisfying the modified restricted eigenvalue condition. As shown in the simulation study reported in \cite{Bogdan2021}, the superior predictive properties of SLOPE are even more pronounced under strongly correlated designs, which is in accordance with the theoretical results of \cite{Figueiredo2014}.

More recently, in \cite{LeeSobczykBogdan2019}, SLOPE has been applied to Gaussian graphical models. Similarly as in the Neighborhood Selection version of the graphical LASSO \citep{Meinshausen2006}, the method relies on application of SLOPE for solving the system of multiple regression problems (\ref{eq:localreg}) and has been given a name the Neighborhood Selection Sorted L-One Penalized Estimator (nsSLOPE). In this article we follow the path of the development of Glasso and propose a novel Gslope algorithm, where the Sorted L-One norm is directly applied to penalize the multivariate normal likelihood.
We propose the selection of the tuning parameters which provably controls the probability of connecting the disjoint components of the graph and yields the procedure which is less conservative than the corresponding version of Glasso (see \cite{Banerjee2008}). Moreover, we propose an even more liberal sequence, which, according to the empirical results, allows to control the False Discovery Rate among the selected edges when the covariance matrix has a block diagonal structure. Furthermore, we extend the approach of \cite{Finegold2011} and construct Tslope for the graphical representation of the t-distributed data. We empirically show that our selection of the tuning parameters still allows for FDR control when the data are t-distributed. We also present empirical results concerning the precision of the estimation of the sparse covariance matrix and the application of our methods for identifying the gene network structure based on the gene expression data. 
Implementation of our methods and codes for the simulation study are available at \url{ https://github.com/Riccardo-Riccobello/Gslope_Tslope_code.git}.

\section{SLOPE for the Gaussian Graphical Models (Gslope)}\label{sec:Gslope}
In this section we formally define the graphical SLOPE (Gslope) and illustrate its properties with respect to control of the number of false edges.

\subsection{Gslope definition}

We assume that our data consist of $n$ independent realizations $x^{(1)},\ldots,x^{(n)}$ of the $p$ dimensional random vector $\bfX$ from a 
%zero-mean 
multivariate normal distribution  $\mathcal{N}_p\left(\mu, \bfSigma \right)$. 
%where without loss of generality, we can assume that $\bfmu=\textbf{0}$   \citep[see, among others,][]{JonesWest2005, Malioutov2006, Peterson2015, LeeSobczykBogdan2019}. 
Our goal is to infer the graphical representation of $\bfX$; $\mathtt{G}=\left(\mathtt{V},\mathtt{E} \right)$, where the vertices $\mathtt{V}$ correspond to the coordinates of $\bfX$ (variables) and the edges connect those components $X_i$, $X_j$, which are conditionally dependent given all other variables $\bfX_{-(i,j)}\in R^{p-2}$.  

Let us denote by $\bfTheta=\bfSigma^{-1}$ the concentration (or precision) matrix of $\bfX$. It is well known that in this Gaussian graphical model $X_i$ and $X_j$ are conditionally independent if and only if $\bfTheta_{i,j}=0$ \citep{Lauritzen1996,Hastie2017}. Thus our goal reduces to the estimation of $\bfTheta$ and identification of its nonzero elements. This knowledge can be further used to increase the precision of the estimation of ${\bfTheta}$ or $\bfSigma$. 

 In this study, we introduce a new graphical model which builds on the SLOPE (\cite{Bogdan2013,Bogdan2015}) method. In contrast to the neighborhood selection version of graphical SLOPE \citep{LeeSobczykBogdan2019}, mentioned in the Introduction, we penalize the log-likelihood function of our data, similar to the Glasso method \eqref{eq:optimGlasso}. Compared to Glasso, we replace the $\ell_1$-norm with the SL1 penalty on 
 a $p \times p$ precision matrix $\bfTheta$. For this purpose, we first vectorize the upper triangle of $\bfTheta$, creating a new $1 \times m$ vector $\bftheta^\star=\left[\theta^\star_1 \cdots \theta^\star_m \right]=\left[\theta_{1,2} \cdots \theta_{1,p}\; \theta_{2,3} \cdots \theta_{2,p} \cdots \theta_{p-1,p}\right]$, where $m=\frac{p(p-1)}{2}$.\footnote{By using this definition of $\bftheta^\star$, we do not penalize the entries placed on the main diagonal of $\bfTheta$. However, our method could be flexibly generalized to include the entries $\theta_{1,1},\ldots , \theta_{p,p}$.} We then define the following SL1 penalty:
\begin{equation}\label{eq:SL1Gslope}
J_{\bflambda}\left(\bfTheta\right)=\sum_{i=1}^{m}\lambda_{i} \left| \theta^\star \right|_{(i)},
\end{equation}
that, similar to $J_{\bflambda}(\bfbeta)$ in \eqref{eq: 5.29}, sorts the absolute values of the entries of $\bftheta^\star$ in decreasing order, and assigns to each of them a specific tuning parameter, such that $\lambda_{1}\geq...\geq\lambda_m\geq 0$.

Building on the penalty defined in Equation \eqref{eq:SL1Gslope}, Gslope solves the following optimization problem:
\begin{equation}\label{eq:glsopeoptimpprobl}
\widehat{\bfTheta} = \operatorname*{arg\; max}_{\bfTheta >  \textbf{0}}\; \left\lbrace  \log \det \bfTheta  -\tr \left( \bfTheta \bfS  \right) - J_{\bflambda}\left(\bfTheta\right) \right\rbrace\;\;,
\end{equation}  
where $\bfS$ is the sample covariance matrix given in (\ref{eq:samcov}).
%$\bfS=\frac{1}{n}\sum_{j=1}^{n} \bfx^{(j)} \left(https://www.overleaf.com/project/603bc6659d32a0ee87563eec \bfx^{(j)}\right)^\prime$.

 Just like Glasso, Gslope is a convex optimization problem which can be efficiently solved. In Section \ref{sec:admmGlasso} we provide an implementation of the ADMM  (Alternating Direction Method of Multipliers, \cite{Boyd2011}) algorithm  for Gslope, which we developed and applied in our empirical analyses.

\subsection{Control of the number of edges between distinct connectivity components}

\subsubsection{Selection of $\lambda$ for Glasso}
Note that the larger $\bflambda$ sequence, the sparser the solution derived from \eqref{eq:SL1Gslope}, with an increasing number of elements of the precision matrix that tend to vanish. The similar situation occurs for Glasso where different approaches have been proposed to compute the optimal $\lambda$ value. Among them, we mention the cross-validation and BIC-type methods, which are widely used in applied machine learning, as they are flexible and easy to implement, providing at the same time accurate results \citep{Hastie2017}. 

One specific goal in the graphical model estimation is the discovery of as many edges as possible while controlling for the number of falsely detected edges. From the practical perspective, we are mainly interested in controlling the probability of the appearance of false edges between two distinct connectivity components of the true graph. 

For any node $k\in \{1,\ldots,p\}$ let us denote by $C_k$ its connectivity component: the set of all nodes which are connected to the node $k$ through some path in the graph. Moreover, let us denote by $\hat C_k^{\lambda}$ the estimate of $C_k$ obtained by the graphical Lasso with the tuning parameter $\lambda$. 
In \cite{Banerjee2008} the following result is proved.  

\begin{theorem}\label{th:banerjee}
 If the tuning parameter for Glasso is selected as
\begin{equation}\label{eq:banerjee}
\lambda_{\alpha}^{Banerjee} = \max_{i>j}(\sqrt{S_{ii} S_{jj}})\frac{t_{n-2}\left(1-\frac{\alpha}{2p^2}\right)}{\sqrt{n-2+t^{2}_{n-2}\left(1-\frac{\alpha}{2p^{2}}\right)}}\;\;,
\end{equation}
where $S_{ii}$ is the empirical variance of $i$-th variable and $t_{n-2}(\delta)$ is the $\delta$ quantile of the student's t-distribution with $n-2$ degrees of freedom,  then
$$P\left (\forall k\in \{1,\ldots,p\}: \hat C_k^{\lambda} \subset C_k \right)\geq 1-\alpha\;\;.$$
\end{theorem}

\begin{corollary}
According to Theorem \ref{th:banerjee}, the probability of connecting different connectivity components by Glasso with $\lambda=\lambda_{\alpha}^{Banerjee}$ is not larger than $\alpha$.
\end{corollary}

While the proof of Theorem \ref{th:banerjee} is not trivial, the selection of $\lambda$ is motivated by the following basic facts:
\begin{itemize}
    \item Two variables from distinct connectivity components are not correlated.
    \item Let $r$ be the sample correlation coefficient between $X_i$ and $X_j$. If the vector $(X_i,X_j)$ has a bivariate normal distribution and $Cov(X_i,X_j)=0$ then the statistic
    $$t=r\sqrt{\frac{n-2}{1-r^2}}\;\;$$
    has a t-distribution with $n-2$ degrees of freedom.
    \item Bonferroni correction: The probability of at least one false rejection (Family Wise Error Rate, FWER) in the sequence of $v$ tests is smaller than the sum of type I errors for each of these tests. Thus, to control FWER  at the level $\alpha$, one can perform each test at the significance level $\alpha/v$.
\end{itemize}

The above observations lead to the conclusion that the tuning parameter $\lambda_{\alpha}^{Banerjee}$ is actually too conservative (i.e. too large). This is because the construction uses the Bonferroni correction to adjust
to $p^2$ tests, while in fact we test only for $\frac{p(p-1)}{2}$  off-diagonal elements of the precision matrix. The following result states that indeed, the probability of falsely connecting distinct connectivity components can be controlled by Glasso with a substantially smaller tuning parameter $\lambda$.

\begin{theorem}\label{th:Glasso}
 If the tuning parameter for Glasso is selected as
\begin{equation}\label{eq:lamGlasso}
\lambda_{\alpha}^{Bon} = \max_{i>j}(\sqrt{S_{ii} S_{jj}}) \frac{t_{n-2}\left(1-\frac{\alpha}{p(p-1)}\right)}{\sqrt{n-2+t^{2}_{n-2}\left(1-\frac{\alpha}{p(p-1)}\right)}}, 
\end{equation}
where $S_{ii}$ is the empirical variance of $i$-th variable and $t_{n-2}(\delta)$ is the $\delta$ quantile of the student's t-distribution with $n-2$ degrees of freedom, then
$$P\left (\forall k\in \{1,\ldots,p\}: \hat C_k^{\lambda} \subset C_k \right)\geq 1-\alpha\;\;.$$
\end{theorem}

The proof of Theorem \ref{th:Glasso} is presented in Section \ref{Ap1} of the Appendix.

\subsubsection{Controlling the probability of connecting different connectivity components by Gslope}

When applying Gslope we at first standardize our variables to the unit variance. Thus, Gslope is applied to the correlation rather than to the covariance matrix.

As noted above, the tuning parameter $\lambda$ for LASSO provided in Theorem \ref{th:Glasso} is obtained by using the Bonferroni correction for testing the hypotheses about the correlation coefficients. In the theory of multiple testing it is well known that the FWER control can be obtained by using a more liberal Holm procedure, defined below.

\begin{equation}\label{eq:ttest}
\mbox{For}\;\; i\in\{1,\ldots,m\}\;\; \mbox{let}\;\; t_i=\left|r_i\sqrt{\frac{n-2}{1-r_i^2}}\right|
\end{equation}
be the absolute value of the t-test statistic for testing the hypothesis that the two variables ''connected'' by the $i^{th}$ edge are not correlated. Then, let 
$C_j=t_{n-2}\left(1-\frac{\alpha_j}{2}\right)$
be the critical value for the t-test at the significance level $\alpha_j=\frac{\alpha}{m+1-j}$.

Now, let us sort our t-statistics in a non-increasing order $t_{(1)}\geq t_{(2)}\geq \ldots \geq t_{(m)}$ and define
$$k_{min}=\min\;\; \{j:\;\; \mbox{such that}\;\;t_{(j)}< C_j\;\}\;\;.$$
Holm's multiple testing procedure rejects all hypothesis such that $t_i> t_{(k_{min})}$ and controls the probability of rejecting at least one false hypothesis at the level $\alpha$ independently on the structure of correlations between different t-statistics (see \cite{Holm_1979}).

Alternatively, let us now define 
$$k_{max}=\max\;\; \{j:\;\; \mbox{such that}\;\;t_{(j)}\geq C_j\}\;\;.$$
\cite{Hochberg_1988} multiple testing procedure rejects all hypothesis such that $t_i\geq t_{(k_{max})}$. This procedure is more powerful than the Holm's procedure and controls the probability of rejecting at least one false hypothesis at the level $\alpha$ if the t-statistics are independent or satisfy some additional assumptions on their dependency structure, like the multivariate totally positive of order two (MTP2) condition
of \citep{Karlin_1980} or the positive regression dependence on subset (PRDS) \citep{Benjamini_2001, Sarkar_2002}. As discussed in \cite{Karlin_1980} these conditions are shared by commonly encountered multivariate distributions.

The above results give the motivation for the first sequence of the tuning parameters for Gslope:

\begin{gather}\label{Holm_Seq}
\forall k=1, ..., m,\;\;\;\lambda_{k}^{Holm} = \frac{t_{n-2}\left(1-\frac{\alpha}{m+1-k}\right)}{\sqrt{n-2+t_{n-2}^{2}\left(1-\frac{\alpha}{m+1-k}\right)}}\;.
\end{gather}
The following result states that Gslope with the sequence of tuning parameters given in (\ref{Holm_Seq}) controls the probability of connecting disjoint connectivity components at the level $\alpha$. 

\begin{theorem}\label{thm:Gslope_fwer_block}
Assume that the t-test statistics (\ref{eq:ttest}) for testing the hypothesis of the lack of correlation between pairs of variables satisfy the assumptions for the FWER control of the Hochberg's multiple testing procedure. Then it holds
$$P\left (\forall k\in \{1,\ldots,p\}: \hat C_k^{Holm} \subset C_k \right)\geq 1-\alpha\;\;,$$
where $\hat C_k^{Holm}$ are the estimates of the connectivity components  obtained by Gslope with the sequence of the tuning parameters \eqref{Holm_Seq}. 
\end{theorem}

The proof of Theorem \ref{thm:Gslope_fwer_block} is provided in Section \ref{Ap2} of the Appendix.

\begin{figure}
    \centering
    \includegraphics[scale=0.7]{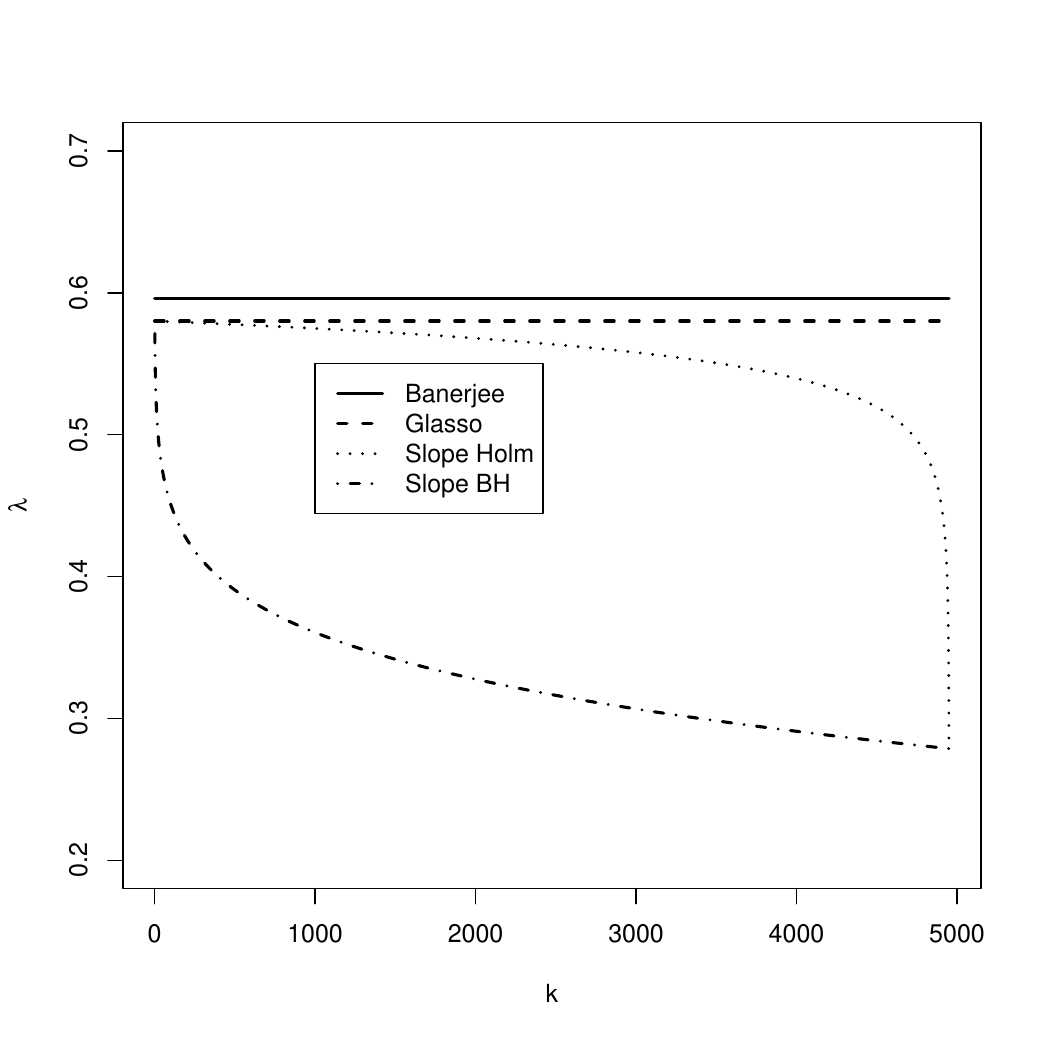}
    \caption{Illustration of the Holm (\ref{Holm_Seq}) and BH (\ref{BH_Seq}) tuning sequences for Gslope for $p=100 $ (i.e., $m=4950$) and $n=50$. The horizontal lines represent the tuning parameters $\lambda$ for Glasso, provided in (\ref{eq:banerjee}) and (\ref{eq:lamGlasso}).}
    \label{fig:lambda}
\end{figure}

\subsubsection{Controlling the distant False Discovery Rate by Gslope}

\begin{figure}
    \centering
    \begin{adjustbox}{width=\textwidth}
    \includegraphics[scale=1]{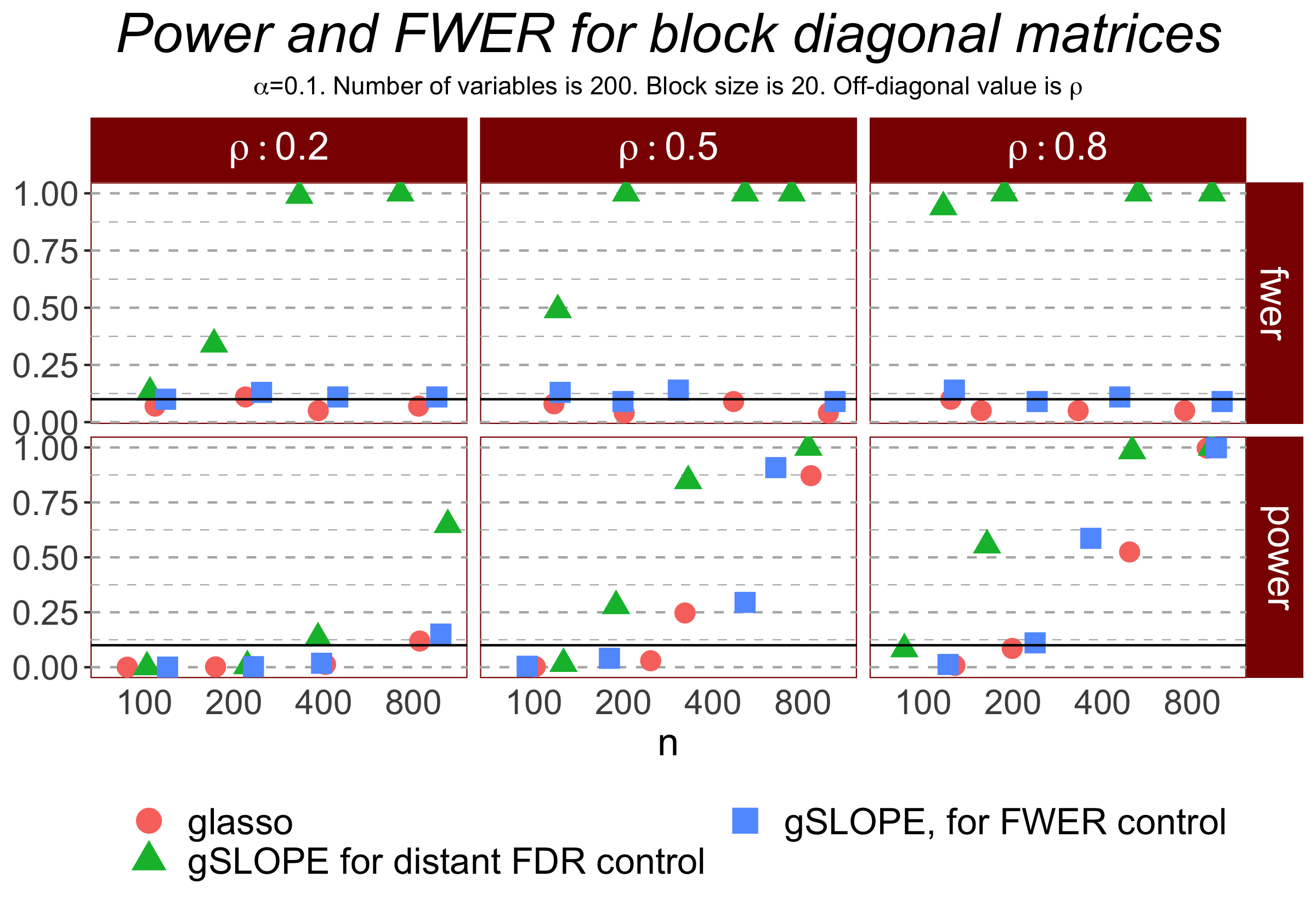} 
    \end{adjustbox}
    \caption{FWER Control. The Figure shows the Family-wise Error Rate (FWER) control for the block diagonal correlation matrix at a given level $\alpha=0.1$ and when considering $n=100, 200, 400, 800$. }
    \label{fig:fwer}
\end{figure}

Figure \ref{fig:lambda} illustrates that for small and moderate $k$ the difference between $\lambda^{Glasso}$ and the elements of the Gslope Holm  sequence is rather small. Therefore, as shown in the Figure \ref{fig:fwer}, the power of identification of important egdes by Gslope is only slightly larger than the respective power of Glasso. 

Therefore, we will now consider a different sequence of the tuning parameters for Gslope, based on the Benjamini-Hochberg correction for multiple testing \citep{Benjamini1995}:
\begin{gather}\label{BH_Seq}
\forall k=1, ..., m,\;\;\;\lambda_{k}^{BH} = \frac{t_{n-2}\left(1-\frac{\alpha k}{2m}\right)}{\sqrt{n-2+t_{n-2}^{2}\left(1-\frac{\alpha k}{2m}\right)}}\;.
\end{gather}

In the context of multiple regression the analogous sequence has been shown to control the proportion of false discoveries among all discoveries (False Discovery Rate, FDR) when the columns of the design matrix are orthogonal.
In \cite{Kos2019,Kos2020} it is proved that the FDR control holds asymptotically, as long as the covariates are independent. In case of the graphical model, the precise control of the False Discovery Rate among edges from the same connectivity component is a very challenging task. However, based on the asymptotic results for the generalizations of Slope (like the logistic regression) presented in \cite{Kos2019} (see also \cite{Kos2020}) we expect that the Gslope based on the BH sequence (\ref{BH_Seq}) should asymptotically control the ''distant'' FDR (dFDR) defined as follows. 

Let $V_d$ be the number of falsely identified edges between different connectivity components and let $R$ be the total number of identified edges. We define dFDR as
$$dFDR=E\left(\frac{V_d}{\max(R,1)}\right)\;\;.$$

\begin{figure}
    \centering
    \begin{adjustbox}{width=\textwidth}
    \includegraphics[scale=1]{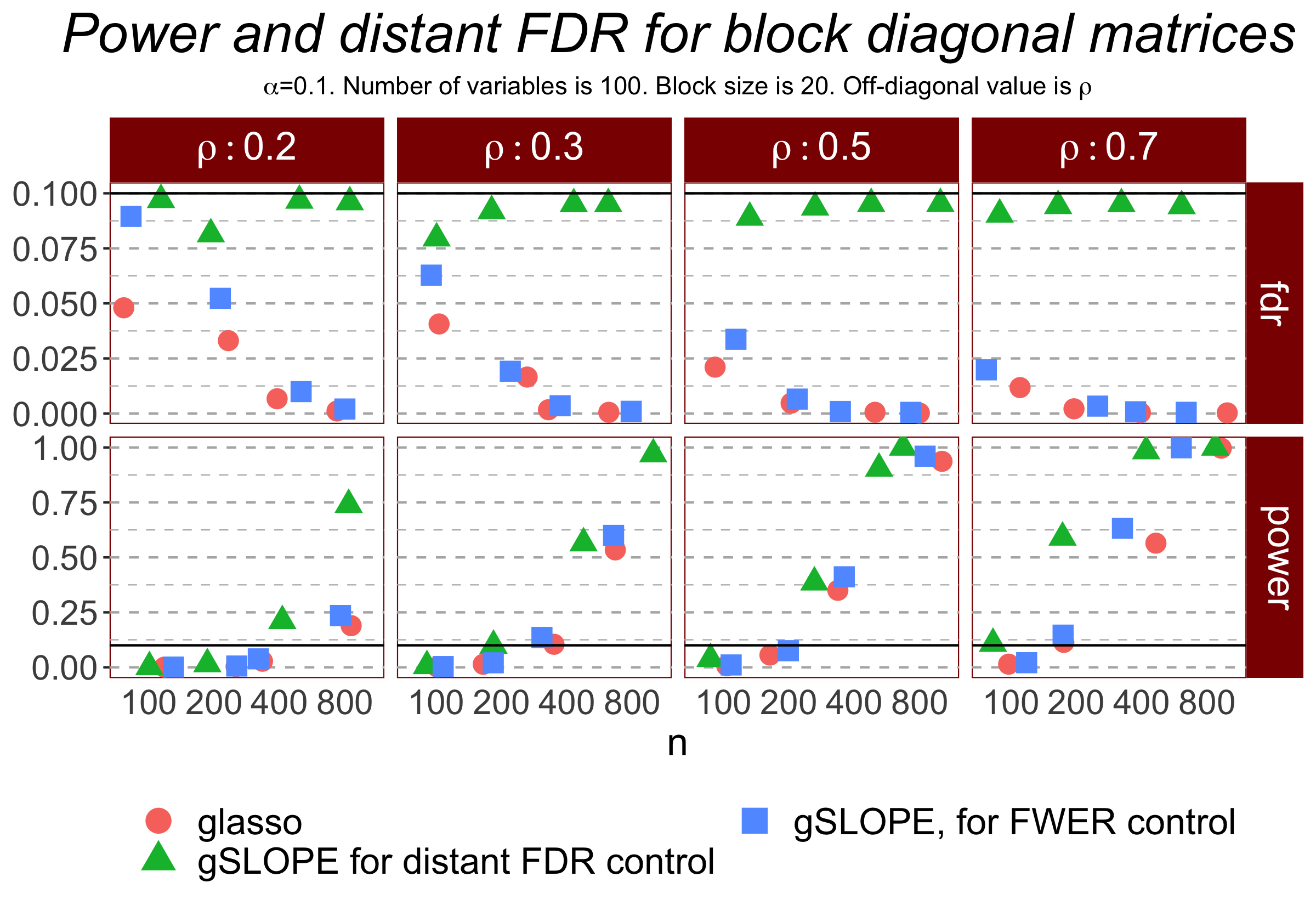} 
    \end{adjustbox}
    \caption{Distant FDR Control. The Figure shows the Distant False Discovery Rate (FDR) control for the block diagonal correlation matrix at a given level $\alpha=0.1$ and when considering $n=100, 200, 400, 800$.}
    \label{fig:distant-fdr}
\end{figure}

The desired performance of the Gslope based on the BH sequence (\ref{BH_Seq}) is shown in Figure \ref{fig:distant-fdr}. The upper panel represents the distant FDR as a function of a sample size for different pairwise correlations between the variable belonging to the same connectivity component. The lower panel represents the power defined as the average percentage of true edges which are identified by different methods. Here, we see that in the examples considered, dFDR is controlled at the assumed level and that Gslope based on the BH sequence can identify many more true edges than Glasso or the Holm version of Gslope. In Section 4.4, we will show that this performance results in improved estimates of the covariance matrix $\Sigma$.

%%% FWER Control - End %%%%%
%

%
%%% Distant FDR Control - Start %%%%%

\subsection{Alternating Direction Method of Multipliers for Gslope}\label{sec:admmGlasso}

The solution to \eqref{eq:glsopeoptimpprobl} can be found using the general alternative direction method of multipliers (ADMM), which has been successfully applied to solve the Glasso optimization problem (see e.g.,  \cite{Scheinberg2010,Boyd2011}). The general formulation of ADMM algorithm can be found in  Appendix \ref{sec:ADMM}.

To derive our algorithm, we first note that the optimization problem in \eqref{eq:glsopeoptimpprobl} is strictly concave. We then derive the corresponding strictly convex and constrained version as follows:
%Problem (\ref{eq: 5.9}) is strictly concave, its convex version is:
\begin{equation}\label{eq:convoptim1}
\begin{cases}
\min_{\bfTheta} & -\log \det \bfTheta + \tr\left(\bfTheta \boldsymbol{S}\right)+J_{\bflambda}\left(\bfTheta\right)\\
s.t. & \bfTheta > \textbf{0}\;\;.
\end{cases},
\end{equation}

%This is a strictly convex constrained optimization problem. Indeed the variable must be a positive semidefinite cone. 
We can rewrite the problem in \eqref{eq:convoptim1} as:
 
\begin{equation}\label{eq:convoptim2}
\begin{cases}
\min_{\bfTheta} & -\log \det \bfTheta +\tr\left(\bfTheta \boldsymbol{S}\right)+\mathbb{I}[\bfTheta >\boldsymbol{0}]+J_{\bflambda}\left(\boldsymbol{Y}\right)\\
s.t. & \bfTheta =\boldsymbol{Y}
\end{cases},
\end{equation}
where the indicator function $\mathbb{I}[A]=0$ if $A$ holds $\mathbb{I}[A]=\infty$ if $A$ does not hold effectively reduces the domain to the positive-definite matrices.
%and  is an indicator function, that we employ to stress the fact that the problem in \eqref{eq:convoptim2} is defined if and only if $\bfTheta$ is a semidefinite matrix, }

Furthermore, the augmented Lagrangian function in the inner product form of the problem in \eqref{eq:convoptim2} is expressed as:
\begin{eqnarray}\label{eq:ALINNPRDCT}
\mathcal{L}^{+}(\bfTheta ,\boldsymbol{Y},\boldsymbol{Z}) & = & -\log \det \bfTheta +\tr\left( \bfTheta \boldsymbol{S}\right)+\mathbb{I}[\bfTheta >\boldsymbol{0}]+J_{\bflambda}\left(\boldsymbol{Y}\right) \\
 &  & +\rho\left\langle \boldsymbol{Z},\bfTheta -\boldsymbol{Y}\right\rangle _{F}+\frac{\rho}{2} \left\Vert \bfTheta -\boldsymbol{Y}\right\Vert_{F}^{2}\nonumber, 
\end{eqnarray}
where $\boldsymbol{Z}$ is the second dual variable, $\rho>0$ is the augmented Lagrangian penalty parameter, $\left\langle \cdot,\cdot\right\rangle _{F}$ is the Frobenius
inner product and $\left\Vert\cdot\right\Vert_{F}$ is the Frobenius norm inducted by the inner product.\footnote{The Frobenius
inner product and the Frobenius norm in Equation \eqref{eq:ALINNPRDCT} are defined, respectively, as $\left\langle \boldsymbol{A},\boldsymbol{B}\right\rangle _{F}=\sum_{i,j}a_{i,j}b_{i,j}=\tr\left( \boldsymbol{A}^\prime \boldsymbol{B}\right)$, where $\boldsymbol{A}^\prime$ is the transposition of $\boldsymbol{A}$, and $\left\Vert\boldsymbol{A}\right\Vert_{F}=\sqrt{\left\langle \boldsymbol{A},\boldsymbol{A}\right\rangle _{F}}=\sqrt{\sum_{i,j}\left| a_{i,j}\right|^{2}}$.} 

Following the ADMM algorithm described in Appendix \ref{sec:ADMM}, we minimize the augmented Lagrangian as a function of $\bfTheta$ and $\boldsymbol{Y}$, implementing the dual update. As for the $(k+1)$-th update of $\bfTheta$, we obtain:\footnote{We report the derivation of Equation \eqref{eq:upomega} in Appendix \ref{sec:omegaupdate}.}
\begin{eqnarray}\label{eq:upomega}
\bfTheta_{(k+1)}= \arg\min_{\bfTheta > \textbf{0}} \left\lbrace  -\log \det \bfTheta +\frac{\rho}{2} \left\Vert \bfTheta +(\boldsymbol{Z}_{(k)}-\boldsymbol{Y}_{(k)}+\rho^{-1}\boldsymbol{S})\right\Vert_{F}^{2} \right\rbrace .
\end{eqnarray}

After defining the quantity $\boldsymbol{\tilde{S}}_{(k)} = -\boldsymbol{Z}_{(k)}+\boldsymbol{Y}_{(k)}-\rho^{-1}\boldsymbol{S}$, 
%\[
%\ensuremath{\boldsymbol{\tilde{S}}^{k}} =-\boldsymbol{Z}^{k}+\boldsymbol{Y}^{k}-\rho^{-1}\boldsymbol{S}
%\]
the optimization problem in \eqref{eq:upomega} can be rewritten as:
\begin{equation}\label{eq:auglagr}
\bfTheta_{(k+1)}=\arg\min_{\bfTheta \geq \textbf{0}} \left\lbrace -\log \det \bfTheta +\frac{\rho}{2}\left\Vert\bfTheta-\ensuremath{\boldsymbol{\tilde{S}}_{(k)}}\right\Vert_{F}^{2} \right\rbrace.
\end{equation}

The gradient of the augmented Lagrangian in \eqref{eq:auglagr} is equal to:
\begin{equation}\label{eq:auglagr2}
\nabla_{\bfTheta}\mathcal{L}^{+}(\bfTheta,\boldsymbol{Y}_{(k)},\boldsymbol{Z}_{(k)})=-\bfTheta^{-1}+\rho\bfTheta-\rho\ensuremath{\boldsymbol{\tilde{S}}_{(k)}}
\end{equation}
and, given the convexity of the augmented Lagrangian in \eqref{eq:auglagr2}, for some optimal matrix $\bfTheta^{*}>\boldsymbol{0}$, we have:
$\nabla_{\bfTheta}\mathcal{L}^{+}(\bfTheta^{*},\boldsymbol{Y}_{(k)},\boldsymbol{Z}_{(k)})=\boldsymbol{0}$, so that:
\begin{equation}\label{eq:optimcondomega}
-(\bfTheta^{*})^{-1}+\rho\bfTheta^{*}-\rho\ensuremath{\boldsymbol{\tilde{S}}_{(k)}}=\boldsymbol{0}.
\end{equation}

This means that $\bfTheta^{*}$ is the solution for the update of $\bfTheta$. We need to find a positive definite solution $\bfTheta^{*}$ to guarantee the invertibility of the precision matrix. We start from the spectral decomposition of $\ensuremath{\boldsymbol{\tilde{S}}_{(k)}}$, defined as $\boldsymbol{\tilde{S}}_{(k)}=\boldsymbol{Q} \bfH \boldsymbol{Q}^\prime$, where $\bfH=\diag\{h_{i}\}$. Suppose that there exists a solution of the following type: $\bfTheta^{*}=\boldsymbol{Q}\boldsymbol{D}\boldsymbol{Q}^\prime$, where $\boldsymbol{D}=\diag\{d_{i}\}$ and $d_{i}>0$. Building on these decompositions, we can reformulate the optimal conditions in \eqref{eq:optimcondomega} as $\boldsymbol{Q}(-\boldsymbol{D}^{-1}+\rho\boldsymbol{D}-\rho \bfH)\boldsymbol{Q}^\prime=\boldsymbol{0}$,
%\begin{equation}
%\boldsymbol{Q}(-\boldsymbol{D}^{-1}+\rho\boldsymbol{D}-\rho\boldsymbol{\Lambda})\boldsymbol{Q}^{T}=\boldsymbol{0},
%\end{equation}
which is equivalent to solve the following equality: 
%Solving equation (\ref{eq: 5.24}) is equivalent to solve:
\begin{equation}\label{eq:optequconds}
-\boldsymbol{D}^{-1}+\rho\boldsymbol{D}-\rho \bfH =\boldsymbol{0}.
\end{equation}

Moreover, given that both $\boldsymbol{D}$ and $\bfH$ are diagonal matrices, Equation \eqref{eq:optequconds} can be expressed as $-d_{i}^{-1}+\rho d_{i}-\rho h_{i}=0 \;\forall i$, from which we obtain $d_{i}^{2}-d_i h_{i}-\frac{1}{\rho}=0$, which leads to the following solution:
%\[
%-d_{i}^{-1}+\rho d_{i}-\rho\lambda_{i}=0,\quad\forall i
%\]
%We can equivalently solve:
%\[
%d_{i}^{2}-\lambda_{i}-\frac{1}{\rho}=0
%\]
%Which gives the following trivial solution (considering only $d_{i}\geq0$):
\begin{equation}
d_{i}=\frac{1}{2}\cdot\left\{ h_{i}+\sqrt{h_{i}^{2}+\frac{4}{\rho}}\right\}. 
\end{equation}

We stress the fact that all diagonal elements are positive, given that $\rho>0$. Moreover, $d_{i}>0$ even if $h_{i}\leq0$. As a result, the condition that $\ensuremath{\boldsymbol{\tilde{S}}_{(k)}}$ has positive eigenvalues is not required. 
%we don't need $\ensuremath{\boldsymbol{\tilde{S}}^{k}}$ to have non-negative eigenvalues. 
Therefore, $\bfTheta^{*}=\boldsymbol{Q}\boldsymbol{D}\boldsymbol{Q}^\prime$ is the solution to our problem, where $\boldsymbol{D}=1/2\cdot \diag\left\{ h_{i}+\sqrt{h_{i}^{2}+\frac{4}{\rho}}\right\} $. Since the solution depends on the eigenvalues of $\ensuremath{\boldsymbol{\tilde{S}}_{(k)}}$, the update rule for $\bfTheta_{(k+1)}$ can be defined as follows:
\begin{equation}\label{eq:updatedruleomega}
\bfTheta_{(k+1)}=\mathcal{F}_{\rho}\left(\ensuremath{\boldsymbol{\tilde{S}}_{(k)}}\right)=\mathcal{F}_{\rho}\left(-\boldsymbol{Z}_{(k)}+\boldsymbol{Y}_{(k)}-\frac{1}{\rho}\boldsymbol{S}\right),
\end{equation}
where $\mathcal{F}_{\rho}\left(\ensuremath{\boldsymbol{\tilde{S}}_{(k)}}\right)\equiv\boldsymbol{Q}\boldsymbol{D}\boldsymbol{Q}^\prime$. 

As for the update rule of $\boldsymbol{Y}_{(k+1)}$, we obtain the following result:
\begin{eqnarray}\label{eq:updateYk1}
\boldsymbol{Y}_{(k+1)}= & \arg\min_{\boldsymbol{Y}} & \mathcal{L}^{+}\left(\bfTheta_{(k+1)},\boldsymbol{Y},\boldsymbol{Z}_{(k)}\right)\nonumber \\
= & \arg\min_{\boldsymbol{Y}} & -\log\det\bfTheta_{(k+1)}+\tr\left(\bfTheta_{(k+1)}\boldsymbol{S}\right)+J_{\bflambda}\left(\boldsymbol{Y}\right)+\rho\left\langle \boldsymbol{Z}_{(k)},\bfTheta_{(k+1)}-\boldsymbol{Y}\right\rangle _{F}\nonumber \\
 &  & +\frac{\rho}{2}\left\Vert\bfTheta_{(k+1)}-\boldsymbol{Y}\right\Vert_{F}^{2}\nonumber \\
= & \arg\min_{\boldsymbol{Y}} &J_{\bflambda}\left(\boldsymbol{Y}\right) +\frac{\rho}{2}\left\Vert\boldsymbol{Y}-(\bfTheta_{(k+1)}+\boldsymbol{Z}_{(k)})\right\Vert_{F}^{2}
=  \mathrm{prox}_{(J_{\lambda},\rho)}(\bfTheta_{(k+1)}+\boldsymbol{Z}_{(k)}),
\end{eqnarray}
where 
$$\mathrm{prox}_{(J_{\lambda},\rho)}(t)=\arg \min_x J_{\lambda}(x)+\frac{\rho}{2}||t-x||_F^2$$
is the proximal operator of the SLOPE norm. An efficient algorithm for solving this proximal optimization problem is provided e.g. in \cite{Bogdan2015}.

Building on the results derived above, we report below the ADMM algorithm that we developed for solving the Gslope problem.

\begin{algorithm}[H]\label{alg:admm_gslope}
\SetAlgoLined

$\boldsymbol{Y}_{(0)}\leftarrow\boldsymbol{\tilde{Y}}$,\:$\boldsymbol{Z}_{(0)}\leftarrow\boldsymbol{\tilde{Z}}$,\:$\rho\leftarrow\rho_{(0)}>0$,\:$k\leftarrow1$\;
	
	\While{convergence criterion is not satisfied}
		{Perform the spectral decomposition of the $\ensuremath{\boldsymbol{\tilde{S}}_{(k)}}$ matrix: $\ensuremath{\boldsymbol{\tilde{S}}_{(k)}}=\boldsymbol{Q}\bfH\boldsymbol{Q}^\prime$\;
		$\bfTheta_{(k+1)}\coloneqq\mathcal{F}_{\rho}(\ensuremath{\boldsymbol{\tilde{S}}_{(k)}})$\;  
		$\boldsymbol{Y}_{(k+1)}\coloneqq \mathrm{prox}_{(J_{\lambda},\rho)}(\bfTheta_{(k+1)}+\boldsymbol{Z}_{(k)})$\; 
		$\boldsymbol{Z}_{(k+1)}\coloneqq\boldsymbol{Z}_{(k)}+\rho(\bfTheta_{(k+1)}-\boldsymbol{Y}_{(k+1)})$\;
		$k \leftarrow k+1$\;}

\caption{Alternating Direction Method of Multipliers for Gslope} \end{algorithm}

\noindent

\begin{theorem}\label{th:theorem}
Algorithm (\ref{alg:admm_gslope}) produces a sequence of precision matrices which converges to the optimal
solution in the objective function value. 
\end{theorem}

\begin{proof}
~\\
 
\begin{itemize}
    \item 
Firstly, notice that both functions that are part of the objective function,
$-\log \det \bfTheta + \tr\left(\bfTheta \boldsymbol{S}\right) +\mathbb{I}[\bfTheta > \boldsymbol{0}]$
and
$J_{\bflambda}\left(\bfTheta\right)$,
 are closed and proper convex functions.
 \item Secondly, by the Saddle Point Theorem (see \citep{Boyd:2004:CO:993483}),  the unaugmented Lagrangian of the Gslope has a saddle point.
 \end{itemize}
\hfill 
Thus, the assumptions for the convergence of the ADMM algorithm from \citep{Boyd2011} are met, which concludes our proof.
\end{proof}

\section{Dealing with $t$-distributed data}\label{sec:Tslope}

The methods discussed so far rely on the assumption that $\bfX$ follows a multivariate normal distribution. However, there exist contexts in which this assumption is restrictive. Therefore, the accuracy of the resulting estimates and inference may be undermined by potential deviations from Gaussianity. For instance, the impact of false-positive edges in Glasso-estimated graphs significantly increases in the presence of heavy-tailed distributions \citep{Cribben2019}. %The Gaussian assumption is then restrictive in many applications, preventing us to properly process the information conveyed by rare but relevant phenomena. 
This limitation has prompted the development of alternative methods. Among them, the Tlasso method introduced by \cite{Finegold2011} has proven to be an effective tool for a robust graphical inference in presence of outliers or contaminated data \citep{Finegold2014, TORRI201851, Cribben2019, TorriGiacomettiPaterlini2019}. In this section, we extend the method introduced by \cite{Finegold2011} with a new Tslope algorithm, which exploits the advantages of the SL1 penalty described in Section \ref{sec:Gslope}.  

In contrast to Section 2, we now assume that $\bfX$ follows a multivariate $t$-distribution with $\nu>2$ degrees of freedom; that is, $\bfX \sim t_p \left( \bfmu, \bfPsi , \nu  \right)$, where 
%$\bfmu$ and $\bfPsi$ are the first two moments of the $t$-distribution. Specifically, 
$\bfmu = \mathbb{E}\left[\bfX  \right] \in \mathbb{R}^p$ is the expected value vector of $\bfX$, whereas $\bfPsi$ is the $p \times p$ positive definite dispersion matrix.  Under the assumption $\bfX \sim t_p \left( \bfmu, \bfPsi , \nu  \right)$, the density function of $\bfX$ is defined as follows:
\begin{equation}\label{eq:densitymt}
f_{\bfX}\left(x \mid \bfmu, \bfPsi, \nu \right)
=\frac{\Gamma[(\nu+p)/2]}{(\pi \nu)^{\frac{p}{2}}\Gamma(\nu/2) \left(\det \bfPsi \right)^{\frac{1}{2}}}
\left[1+\frac{1}{\nu} \left(\bfx - \bfmu \right)^\prime \bfPsi^{-1} \left(\bfx - \bfmu \right)\right]^{-\frac{\nu+p}{2}},
\end{equation}
where $\Gamma (\cdot)$ denotes the gamma function, 
%\red{define $\bfx^{(j)}$ here}.
%whereas $\bfx^{(j)}$ is the $j$-th realization of the random vector $\bfX$, for $j=1,\ldots , n$. 

Given $n$ realizations of the random vector $\bfX \sim t_p \left( \bfmu, \bfPsi , \nu  \right)$, denoted as $\bfx^{(1)},\ldots,\bfx^{(n)}$,
%the sample covariance matrix $\bfSigma$ is defined as
%the sample mean vector: $\widehat{\pmb{\mu}}=\frac{1}{n}\sum_{i=1}^{n}\textbf{x}_i$ and 
%: $\bfS= \widehat{\bfSigma}=\frac{1}{n}\sum_{j=1}^{n} \bfx^{(j)} \left( \bfx^{(j)}\right)^\prime$.
the log-likelihood of the data takes the following form:
%from which we compute the corresponding log-likelihood function:
\begin{eqnarray}\label{eq:loglikmt}
\mathcal{L}\left( \bfmu, \bfPsi, \nu \right) &=& \log\prod_{j=1}^{n} f_{\bfX}\left(\bfx^{(j)} \mid \bfmu, \bfPsi, \nu \right) \\ &=& 
-n\log\Gamma\left(\frac{\nu}{2}\right)+n\log\Gamma\left(\frac{\nu+p}{2}\right)-\frac{np}{2}\log \nu-\frac{np}{2}\log\pi 
 \nonumber \\ &-& \frac{n}{2}\log \det \bfPsi^{-1}   -\left(\frac{\nu+p}{2}\right)\sum_{j=1}^{n}\log\left[\nu+\delta_{\bfx^{(j)}}\left(\bfmu, \bfPsi \right)\right], 
\end{eqnarray}
where $\delta_{\bfx^{(j)}}\left(\bfmu,\bfPsi \right)=\left(\bfx^{(j)} -\bfmu \right)^\prime \bfPsi^{-1} \left(\bfx^{(j)} -\bfmu \right)$ is the the Mahalanobis distance.

It is important to highlight the fact that the covariance matrix of $\bfX$ (i.e. $\bfSigma$) can be expressed as a function of $\bfPsi$,  
%Interestingly, the covariance matrix of $\bfX$ (i.e. $\bfSigma$) 
through the following relationship: $\bfSigma = \frac{\nu}{\nu -2} \bfPsi$. Following \cite{Finegold2011}, for notational convenience, we define the precision matrix of a multivariate $t$-distribution as $\bfTheta=\bfPsi^{-1}$, so that we have a clear connection with the Gaussian graphical model described in Sections 1 and 2.  
%Following \cite{Finegold2011}, we point out the connection with Gaussian graphical models by defining $\bfTheta=\bfPsi^{-1}$.  
%The precision matrix will be then given by: $\boldsymbol{\Sigma}^{-1}=\frac{v-2}{v}\boldsymbol{\Phi}^{-1}$, for notational convenience we denote the ``precision matrix'' of a $t$-student distribution as in the gaussian case: $\boldsymbol{\Phi}^{-1}\equiv\boldsymbol{\Omega}$, despite we know that the precision matrix is given by $\boldsymbol{\Sigma}^{-1}$.
Similar to the Glasso and Gslope methods (see Sections 1 and 2), we aim at estimating networks in which vertices $i$ and $j$ are not connected by an edge if $\theta_{i,j}=0$. We still employ a penalty function to promote sparsity in the precision matrix. Nevertheless, in contrast to the Gaussian framework, the condition $\theta_{i,j}=0$ no longer implies conditional independence when using the $t$-distribution \citep{Baba2004}. However, despite the lack of conditional independence, the $t$-distribution provides the following property: if two vertices $i$ and $j$ are separated by a set of nodes $\mathtt{C}$ in $\mathtt{G}$, then $X_i$ and $X_j$ are conditionally uncorrelated given $X_{\mathtt{C}}$ \citep{Finegold2011}. Therefore, it is reasonable to replace conditional independence with zero partial correlation or zero conditional correlation. By doing so, disconnected nodes in a given graph can be considered orthogonal to each other after the effects of other vertices of the same network are removed \citep{TORRI201851}. 
%As highlighted by \cite{Finegold2011}, zero conditional correlations implies that a mean-square error optimal prediction of the variable $X_i$ can be based on the neighbors of the node $i$ in the graph.

When assuming $\bfX \sim t_p \left( \bfmu, \bfPsi , \nu \right)$, the lack of density factorization properties with $t$-distributions complicates the likelihood inference \citep{Finegold2011}. However, we can efficiently estimate the parameters of interest by implementing the Expectation-Maximization (EM) algorithm proposed by \cite{Finegold2011}, that we adapt to our Tslope specification. In particular, the EM algorithm builds on the scale-mixture representation of the $t$-distribution \citep{Finegold2011} described below, which leads to relevant computational advantages.
%In this context, it is useful In order to connect the \textit{Graphical Slope} with the T-student specification, we can to consider a scale-mixture representation
%\footnote{As argued in \citet{49}, the scale-mixture representation clarifies how the use of $t$-student distribution leads to more robust inference, as extreme observations can arise from small values of $\tau$.} 
%of the $t$-distribution \citep{Finegold2011}. 
Specifically, let $\boldsymbol{W}\sim \mathcal{N}_{p}(\boldsymbol{0},\bfPsi)$ be a multivariate normal random vector independent of the Gamma random variable $\tau\sim\Gamma(\nu/2, \nu/2)$, then:
\begin{equation}\label{eq:mixrepres}
\boldsymbol{X}=\bfmu+\frac{\boldsymbol{W}}{\sqrt{\tau}}\sim t_{p}(\bfmu,\bfPsi, \nu).
\end{equation}

This scale-mixture representation allows for easy sampling and emphasizes the fact that the $t$-distribution leads to more robust inference, as extreme observations can be the result of small $\tau$ values \citep{Finegold2011}.
%Where $\boldsymbol{\mu}$ is a $p$-dimensional mean vector, $\boldsymbol{\Phi}$ is a $p\times p$ scale matrix and $v$ are the degrees of freedom.
Moreover, it allows us to derive the conditional distribution of $\tau$ given $\boldsymbol{X}$ and $\nu$ \footnote{Following \cite{Finegold2011}, we assume that the degrees of freedom ($\nu$) are known. Indeed, the estimation of $\nu$, in addition to that of $\bfmu$ and $\bfPsi$, reduces the local robustness of the corresponding estimators. However, if desired, as pointed out by \cite{Finegold2011}, we can also estimate $\nu$ by employing the method discussed by \cite{LiuRubin1995}.}:
\begin{equation}\label{eq: 5.36}
f_\tau \left(\tau \mid \boldsymbol{X}\right)\sim\Gamma\left(\frac{\nu+p}{2},\frac{\nu+\left(\boldsymbol{X}-\bfmu\right)^\prime \bfPsi^{-1}\left(\boldsymbol{X}-\bfmu\right)}{2}\right).
\end{equation}

%%%%%%%%%%%%%%%%%%%%%%%%%%%%%%%%%%%%%%%%%%%%%%%%%%%%%%%%%%%%%%%%%%%%%%%%%%%%%%%%%%%%%%%%%%%%%%%%%%%%%%%%%%%%%%%%%
%%%%%%%%%%%%%%%%%%%%%%%%%%%%%%%%%%%%%%%%%%%%%%%%%%%%%%%%%%%%%%%%%%%%%%%%%%%%%%%%%%%%%%%%%%%%%%%%%%%%%%%%%%%%%%%%%
%%%%%%%%%%%%%%%%%%%%%%%%%%%%%% EM algorithm for Tslope  %%%%%%%%%%%%%%%%%%%%%%%%%%%%%%%%%%%%%%%%%%%%%%%%%%%%%%%%%
%%%%%%%%%%%%%%%%%%%%%%%%%%%%%%%%%%%%%%%%%%%%%%%%%%%%%%%%%%%%%%%%%%%%%%%%%%%%%%%%%%%%%%%%%%%%%%%%%%%%%%%%%%%%%%%%%
%%%%%%%%%%%%%%%%%%%%%%%%%%%%%%%%%%%%%%%%%%%%%%%%%%%%%%%%%%%%%%%%%%%%%%%%%%%%%%%%%%%%%%%%%%%%%%%%%%%%%%%%%%%%%%%%%
 
Equation \eqref{eq: 5.36} immediately implies that conditional on $\tau$ %as a hidden variable, \cite{Finegold2011} derived the conditional %distribution of $\bfX$ given $\tau$, that is defined %as
: 
\begin{equation}\label{eq: 5.39}
f_{\bfX}\left( \bfx \mid \tau\right)\sim \mathcal{N}_{p}\left(\bfmu,\bfPsi/\tau\right).
\end{equation}

%In the E-step, $\tau$ is treated as a hidden variable, whereas the penalized log-likelihood of the latent Gaussian vector is maximized in the M-step, using the Gslope method. 
%We still take into account $n$ realizations of $\bfX$ (i.e. $\bfx^{(j)}$, for $j=1,\ldots , n$) assuming now that they are drawn from the $t_p \left( \bfmu, \bfPsi , \nu \right)$ distribution. 
Now, suppose that we observe the following sequence of the hidden Gamma-random variables:
\begin{equation}\label{eq: 5.40}
\tau^{(1)},...,\tau^{(n)}\sim\Gamma(\nu/2, \nu/2).
\end{equation}

Equations \eqref{eq: 5.39} and \eqref{eq: 5.40} form a scale-mixture model with the following T-slope penalized complete log-likelihood function \citep{Liu1997}:
\begin{equation}
\mathcal{L}\left(\bfmu,\bfTheta \mid \bfx,\bftau,\nu\right) \propto 
  -\frac{1}{2}\sum_{j=1}^{n}\tau^{(j)}\left(\bfx^{(j)}-\bfmu\right)^\prime \bfTheta \left(\bfx^{(j)}-\bfmu\right) +\frac{n}{2}\log \det \bfTheta - J_{\bflambda}\left(\bfTheta\right)\;\;, \nonumber
\end{equation}
where, $\bfx=(x^{(1)},\ldots,x^{(n)})$, $\bftau=(\tau^{(1)},\ldots,\tau^{(n)})$, and the symbol $\propto$ indicates that irrelevant additive constants
are omitted.

Following \cite{Finegold2011}, we employ the EM algorithm that we adapt to our Tslope method. This algorithm is described below:
\begin{itemize}
\item \textbf{E-step}: we compute the conditional expectation of the penalized complete log-likelihood function given the realization $\bfx^{(j)}$. Since the penalized log-likelihood is a linear function of $\tau$, we only need to compute the conditional expectations of the coordinates of $\tau$, which are equal to:
\begin{equation}
\mathbb{E}\left[\tau^{(j)} \mid \bfx^{(j)}\right]=\frac{\nu+p}{\nu+\left[\delta_{\bfx^{(j)}}\left(\bfmu,\bfPsi \right)\right]}.
\end{equation}

Given the current estimates of $\bfmu$ and $\bfPsi^{-1}$, denoted, respectively, as $\widehat{\bfmu}_{(k)}$ and $\widehat{\bfPsi}^{-1}_{(k)}$, we can compute $\hat{\tau}_{(k+1)}^{(j)}$ at the $(k+1)$-th iteration as:
\begin{equation}
\hat{\tau}_{(k+1)}^{(j)}=\frac{\nu+p}{\nu+ \left[\delta_{\bfx^{(j)}}\left( \widehat{\bfmu}_{(k)},   \widehat{\bfPsi}^{-1}_{(k)} \right)\right]},\quad j=1,...,n.
\end{equation}

\item \textbf{M-step}: we maximize the complete log-likelihood to obtain the parameter estimates at iteration $k+1$:
\begin{equation}
\widehat{\bfmu}_{(k+1)}=\frac{\sum_{j=1}^{n}\hat{\tau}_{(k+1)}^{(j)}\bfx^{(j)}}{\sum_{j=1}^{n}\hat{\tau}_{(k+1)}^{(j)}}
\end{equation}

\begin{equation}
\bfS_{(k+1)}=\frac{1}{n}\sum_{j=1}^{n}\hat{\tau}_{(k+1)}^{(j)}  \left(\bfx^{(j)}- \widehat{\bfmu}_{(k+1)} \right)\left( \bfx^{(j)}- \widehat{\bfmu}_{(k+1)}\right)^\prime .
\end{equation}

We implement the Gslope method (see Section \ref{sec:Gslope}) to obtain the estimate of the precision matrix $\bfPsi_{(k+1)}^{-1}\equiv \bfTheta_{(k+1)}$. We remind the reader that the Gslope method solves the following optimization problem:
$\widehat{\bfTheta}_{(k+1)}=\arg\max_{\bfTheta}\left\{ \log \det \bfTheta  -\tr\left(\bfTheta\boldsymbol{S}_{(k+1)}\right)-J_{\bflambda}(\bfTheta)\right\}$ , where $J_{\bflambda}(\bfTheta)$ is the SL1 penalty defined in Equation \eqref{eq:SL1Gslope}. 
\end{itemize}

We iterate the E- and M-steps until we satisfy the following convergence criterion: $\left\Vert \widehat{\bfTheta}_{(k+1)}-\widehat{\bfTheta}_{(k)}\right\Vert<\epsilon$, where $\epsilon$ is a sufficiently small threshold. 
%The \textit{E-step and M-step }are repeated until the convergence criterion $\parallel\hat{\boldsymbol{\Omega}}_{(k+1)}-\hat{\boldsymbol{\Omega}}_{(k)}\parallel<\epsilon$ is met, where $\epsilon$ is a given small number, I have called this procedure \textit{Tslope} algorithm. Convergence to a stationary point is guaranteed and a local maximum is typically found, despite this the maximized log-likelihood function is not concave, and so one finds oneself in the usual situation of not being able to give any guarantees about having obtained a global maximum \citep{49}.

\section{Simulation study}\label{sec:datasimulation}

In this section we report the results of the simulation study comparing Gslope and Tslope to other state-of-art methods for the estimation of the precision matrix.
\subsection{Simulation set-up}
\begin{figure}[h!]
\begin{adjustbox}{width=\textwidth}
\begin{tabular}{cccc}
    \includegraphics[scale=0.25]{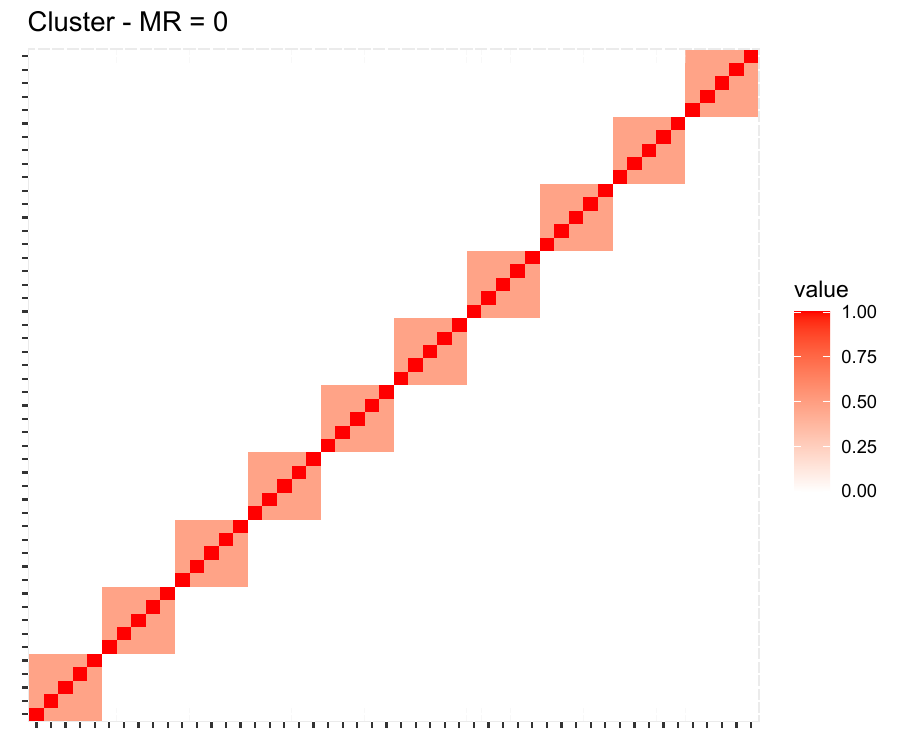} & 
    \includegraphics[scale=0.25]{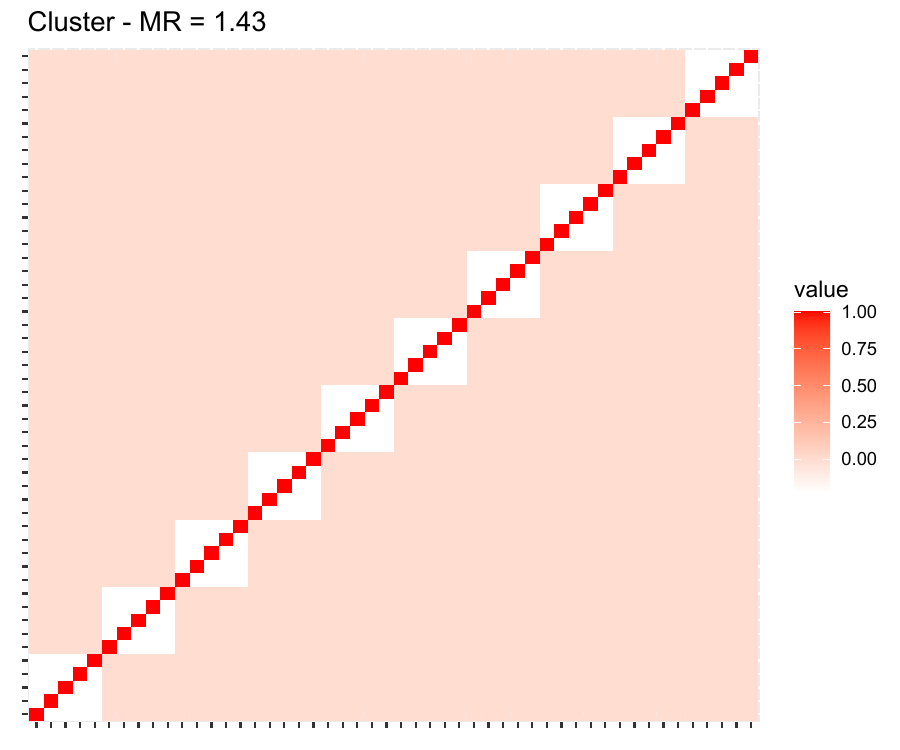} &
    \includegraphics[scale=0.25]{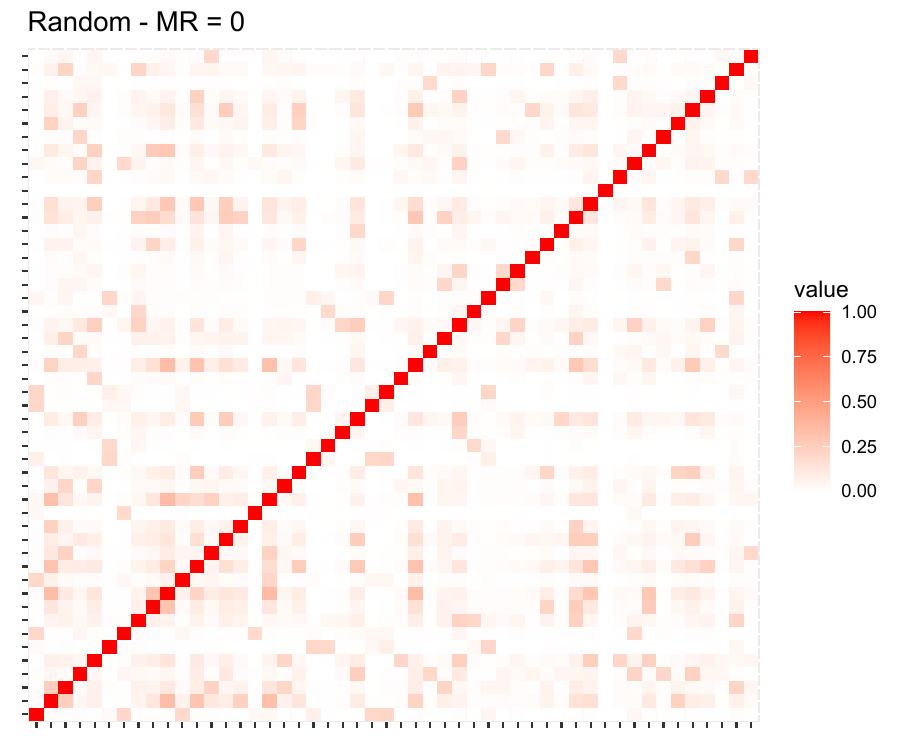} &
    \includegraphics[scale=0.25]{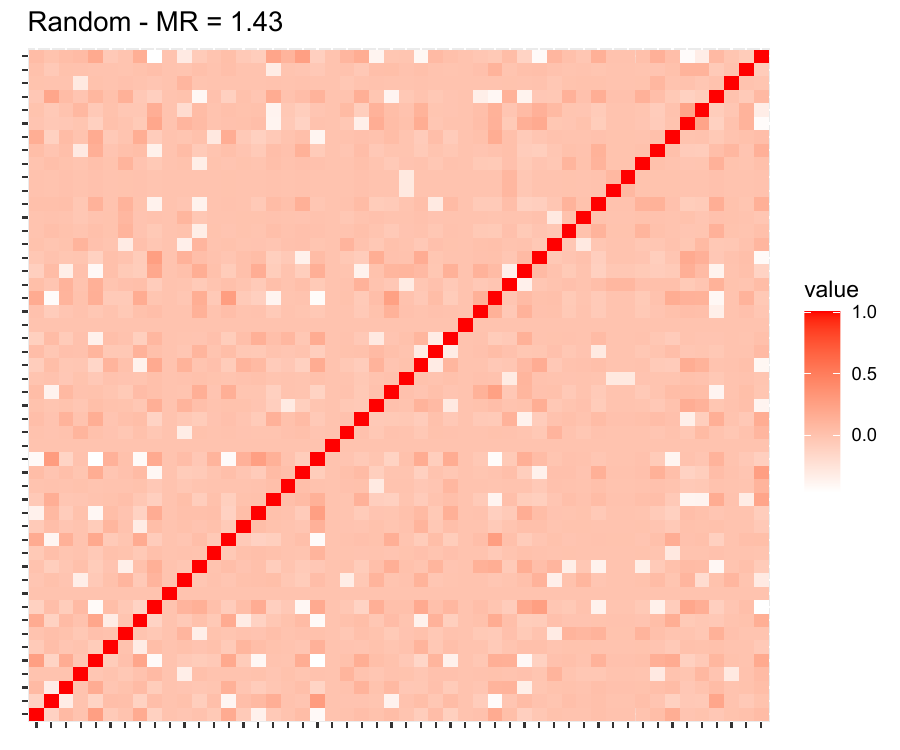}\\
        (a) & (b) & (c) & (d)\\
\end{tabular}
\end{adjustbox}
    \caption{Heatmaps of the partial correlation matrices for cluster networks ((a) and (b)) and the random network structures ((c) and (d)) and the magnitude ratios $MR=0$ and $MR=1.43$.}
    \label{fig:network_structures}
\end{figure}

For our simulation study, we rely on the R package \textit{huge}, which allows to simulate data for undirected graphical models for many different network configurations, including cluster, random, hub, scale-free and band structures.\footnote{For more information on the R package \textit{huge}, please refer to \url{https://CRAN.R-project.org/package=huge}} Having specified the desired network structure, the number of nodes $p$ and the so called \textit{Magnitude Ratio} (MR), the package returns an oracle covariance matrix $\bfSigma$ and an oracle precision matrix, $\bfTheta$. The oracle covariance matrix $\bfSigma$ is then used as an input to a data generating process, from which $n$ data points are sampled. These data are then used to estimate the oracle covariance matrix using the considered methods of estimation of sparse graphical models.\\ 
Here, we resort to presenting the results for the cluster and the random networks, as we believe that those networks capture the majority of real world applications. For example, different stocks can be grouped according to different economic sectors. Still, their relationships are far from perfect and random linkages across such sectors are often observed.\footnote{While we here only focus on the cluster and random network configuration, the results for the other network structures are available on: \url{https://github.com/Riccardo-Riccobello/Gslope_Tslope_code/tree/main/Results_Stats_Paper}.}\\
To set the \textit{Magnitude Ratio} (MR), which is given as $MR = \frac{u+v}{v}$, we need to choose a value for $u$, which is the value that is added to the diagonal elements of the precision matrix after it has been transformed to a positive semi-definite matrix, and $v$ that represent the off-diagonal non-zero entries of the precision matrix. The off-diagonal entries are modified to ensure the invertibility of the covariance matrix. As such, the magnitude ratio regulates the dominance of the diagonal entries, as compared to the off-diagonal entries. If the MR is close to zero, the off-diagonal entries are dominated by the diagonal entries. On the other hand, if MR is large, then the off-diagonal elements dominate. 
%The latter is particularly important when we investigate the \textit{cluster} network structure.
In what follows, we choose MR=0 and MR=1.43 to capture different relationships between the main diagonal and off-diagonal elements.\footnote{We choose these two Magnitude Ratios to evaluate the performance of the algorithms in two distinct settings. Furthermore, our analysis showed that increasing the MR above 1.43 does not lead to more distinct results. More information on the Magnitude Ratio is provided by \cite{ChanWood1997}. } Figure \ref{fig:network_structures} shows the heatmaps of the partial correlation structure of the considered networks. The figure illustrates how a higher magnitude ratio increases the dominance of the off-diagonal elements.\\
Finally, given a sample of $n$ independent $p$-dimensional data vectors, we aim to estimate $\bfTheta$ directly, using our newly introduced Gslope and Tslope algorithm, and further compare its performance to other state-of-the-art methods for estimation the precision matrix, including the inverse of the sample covariance matrix (Sample, only when $n>p$), graphical elastic net (Elastic Net, \cite{Kovacs2021, Bernardini2021}), the Resampling Of Penalized Estimates (ROPE, \cite{Kallus2017}), the Glasso (Glasso, \cite{FriedmanHastieTIbshirani2008}), and the Tlasso (Tlasso, \cite{Finegold2011}).\\
%
%%% Network Structure Figures - Start %%%%
%%% Network Structure Figures - End %%%%
% Distributions
To analyse the robustness of our new methods with regard to the data generating processes, we make the following three assumptions for each of the underlying distributions:
\begin{enumerate}
    \item \textit{Multivariate Gaussian distribution:} $\bfX \sim N_p(0, \bfSigma)$
    \item \textit{Multivariate t-student distribution:} $\bfX \sim t_p(0, \bfSigma, \nu)$, with $\nu=4$
    \item \textit{Multivariate Mixture distribution:} $\bfX \sim 0.5 \times N_p(0, \bfSigma) + 0.5 \times t_p(0, \bfSigma, \nu)$, with $\nu=4$
\end{enumerate}

Given, that we want to compare all methods in situations where both $n>p$ and $n<p$, we fix the number of nodes to be $p=100$, and let the number of observations vary, drawing first $n=250$ and then $n=50$ data points. With the two network configurations (i.e. cluster and random), the two different values of MR (i.e. MR=0 and MR=1.43) and the three different distributional assumptions (i.e. Gaussian, t-student and mixed), we consider a total of 24 distinct simulation set-ups.\\
% Tuning Parameter
\subsection{Tuning parameter set-up}\label{sec:tuningparametersetup}
All methods - except the sample estimate - depend on a single tuning parameter or a decreasing sequence of tuning parameters, which trade off model complexity and sparsity.

For Glasso and Tlasso we choose $\lambda$  according to the formula (\ref{eq:banerjee}) of \cite{Banerjee2008}, with $\alpha=0.05$ for $n=250$ and $\alpha=0.4$ for $n=50$. This goes along the statistical practice of using larger significance levels for smaller sample sizes, which allows to enhance estimation properties by preserving a high power of detection of important edges and reducing the bias due to the $L_1$ norm shrinkage.

\begin{figure}
    \centering
    \begin{adjustbox}{width=\textwidth}
    \begin{tabular}{cc}
    \includegraphics[scale=1]{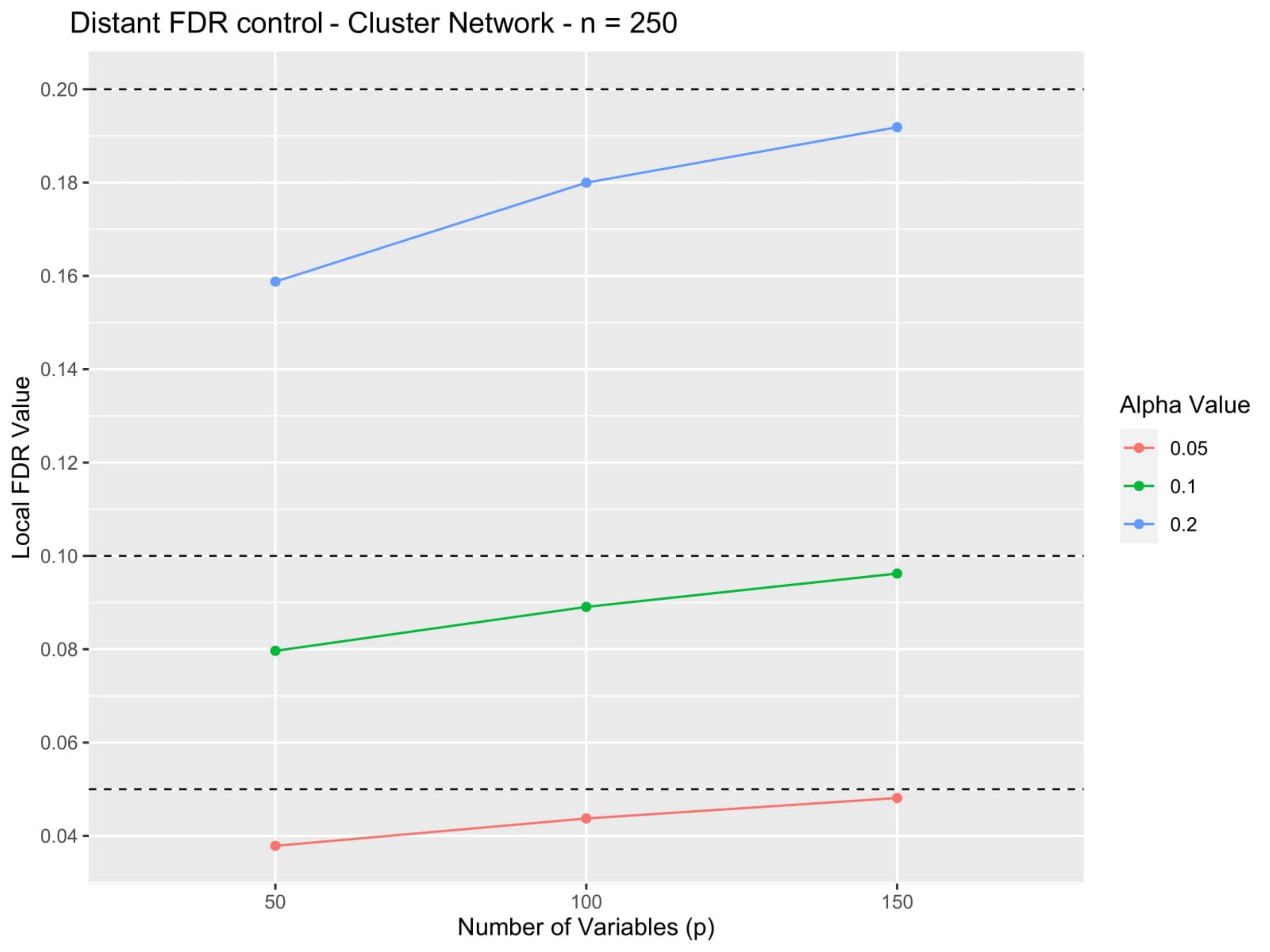} &
    \includegraphics[scale=1]{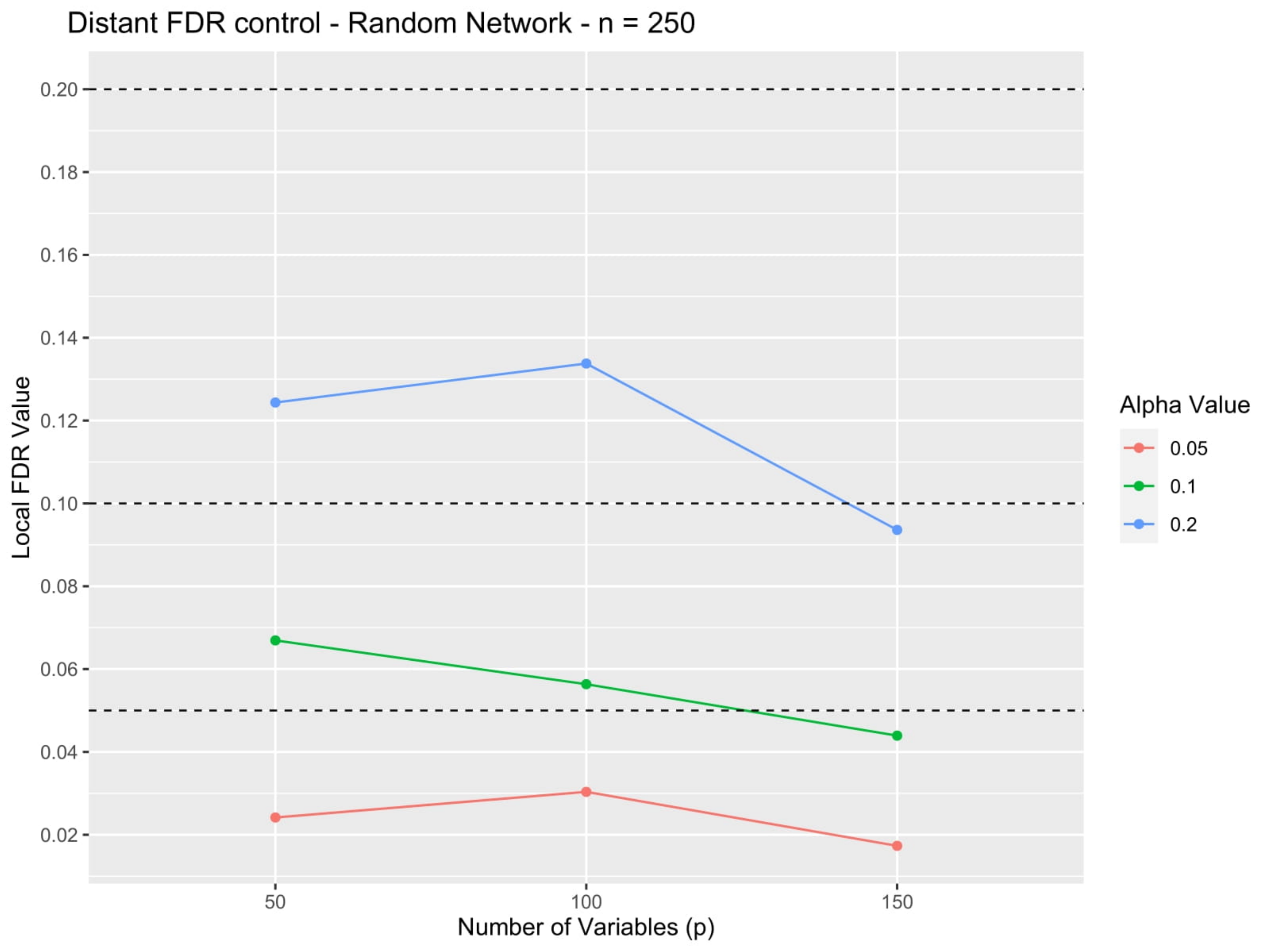}
        \end{tabular}
    \end{adjustbox}
    \caption{The distant False Discovery Rate (FDR) control by Tslope for the data generated according to the multivariate t-distribution and the cluster and random network with $MR=0$. Here $alpha= 0.05, 0.1, 0.2$, $n=250$ and $p=50, 100, 150$, respectively.}
    \label{fig:fdr}
\end{figure}

In case of Gslope and Tslope we resort to the Benjamini-Hochberg (BH) sequence of the tuning paramaters (\ref{BH_Seq}). As illustrated in Section 2, this sequence allows Gslope to control the distant FDR when the data come from the multivariate normal distribution.  Figure \ref{fig:fdr} illustrates that the same sequence allows Tslope to control the distant FDR when the data come from the multivariate t-distribution and $n>p$. For consistency with Glasso, in the remaining part of this section we will set the $\alpha$ parameters for Gslope and Tslope at $\alpha=0.05$ for $n=250$ and $\alpha=0.4$ for $n=50$. As the sequence of tuning parameters for Gslope and Tslope is decaying we expect the overall shrinkage magnitude to be larger for Glasso and Tlasso, as compared to the Slope procedures. Thus, we expect Slope methods to produce denser graphs.

ROPE and Elastic Net require the selection of the tuning parameter for the Lasso part of the algorithm. In our simulations we fixed this parameter at the same value as for Glasso, i.e. according to the formula (\ref{eq:banerjee}). We also verified that the performance of these procedures is not substantially different when the selection of this tuning parameter is performed using cross-validation. Furthermore, for Elastic Net we set the value of the mixing parameter ($\alpha$ in {\it glmnet}) to 0.5 and for ROPE we used the default value of the FDR nominal level $q=0.1$.
%nti could lead to an increased sparsity in the estimated precision matrix.\\
% Number of simulations and performance measures
\subsection{Accuracy characteristics}
While the tuning sequences for Gslope and Tslope were selected for the distant FDR control, it is interesting to verify how they perform with respect to other important measures of the accuracy of the precision matrix estimation. In our study we consider the two following standard accuracy measures.

\begin{itemize}
\item {\bf F1 score.}

\begin{gather}
    %\text{F-Measure} = 2 \frac{|\supp(\bfTheta)| \cap |\supp(\widehat{\bfTheta})|}{|\supp(\bfTheta)| + |\supp(\widehat{\bfTheta})|}
    \text{F1-Score} = \frac{\text{True Positives}}{\text{True Positives} + \frac{1}{2}(\text{False Positives} + \text{False Negatives})}\;\;,
\end{gather}
whereas the True Positives are the number of edges which are both in the true and estimated precision matrix, the False positives are those which are active in the estimated precision matrix, but not in the true one, and the False negatives are those edges, which the procedure fails to identify. 
%whereas the support of a matrix $\bfTheta$ is defined as $\supp(\bfTheta) = \{i,j: \Theta_{i,j} \neq 0\}$.
The F1 statistics is defined on the interval [0,1]. A value of 1 indicates optimal model selection properties, in which only those edges have been selected that are also present in $\bfTheta$.\\

\item {\bf Frobenius Norm.} 
\begin{gather}
D_{F}(\widehat{\bfTheta}, \bfTheta) = ||\widehat{\bfTheta} - \bfTheta|| = \sqrt{(tr((\widehat{\bfTheta}-\bfTheta)(\widehat{\bfTheta} - \bfTheta)')}\;\;,
\end{gather}
where a lower value indicates a higher accuracy of the estimated model. Different to the F1-Score, the Frobenius Norm distance allows us to evaluate the magnitude of the estimated edges as compared to the oracle precision matrix.\\
\end{itemize}

\subsection{Simulation results}
%
%%% F1 Score - n>p - Start %%%%%
\begin{figure}
    \centering
    \begin{adjustbox}{width=\textwidth}
    \begin{tabular}{c}
    \includegraphics[scale=1]{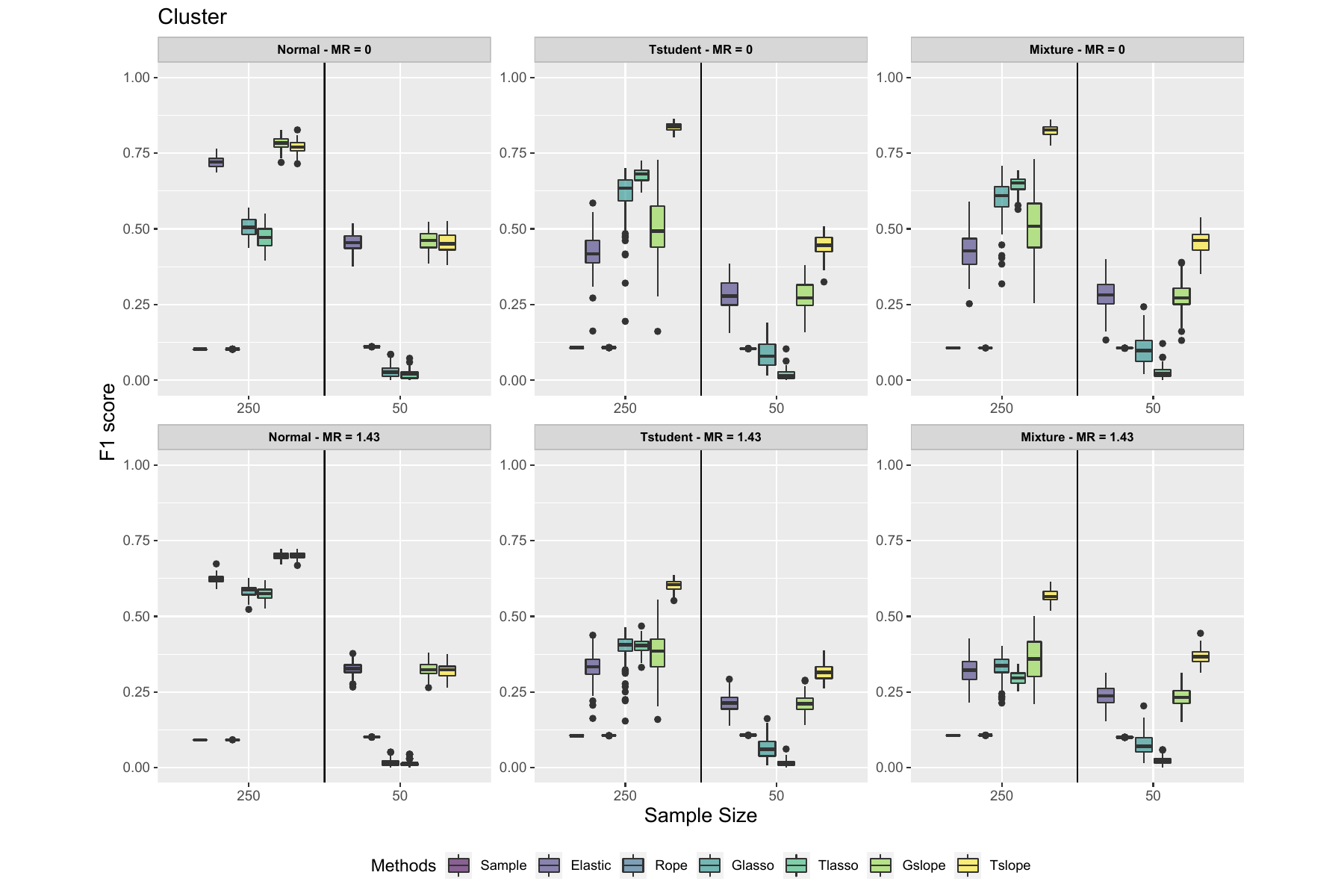} \\
    \includegraphics[scale=1]{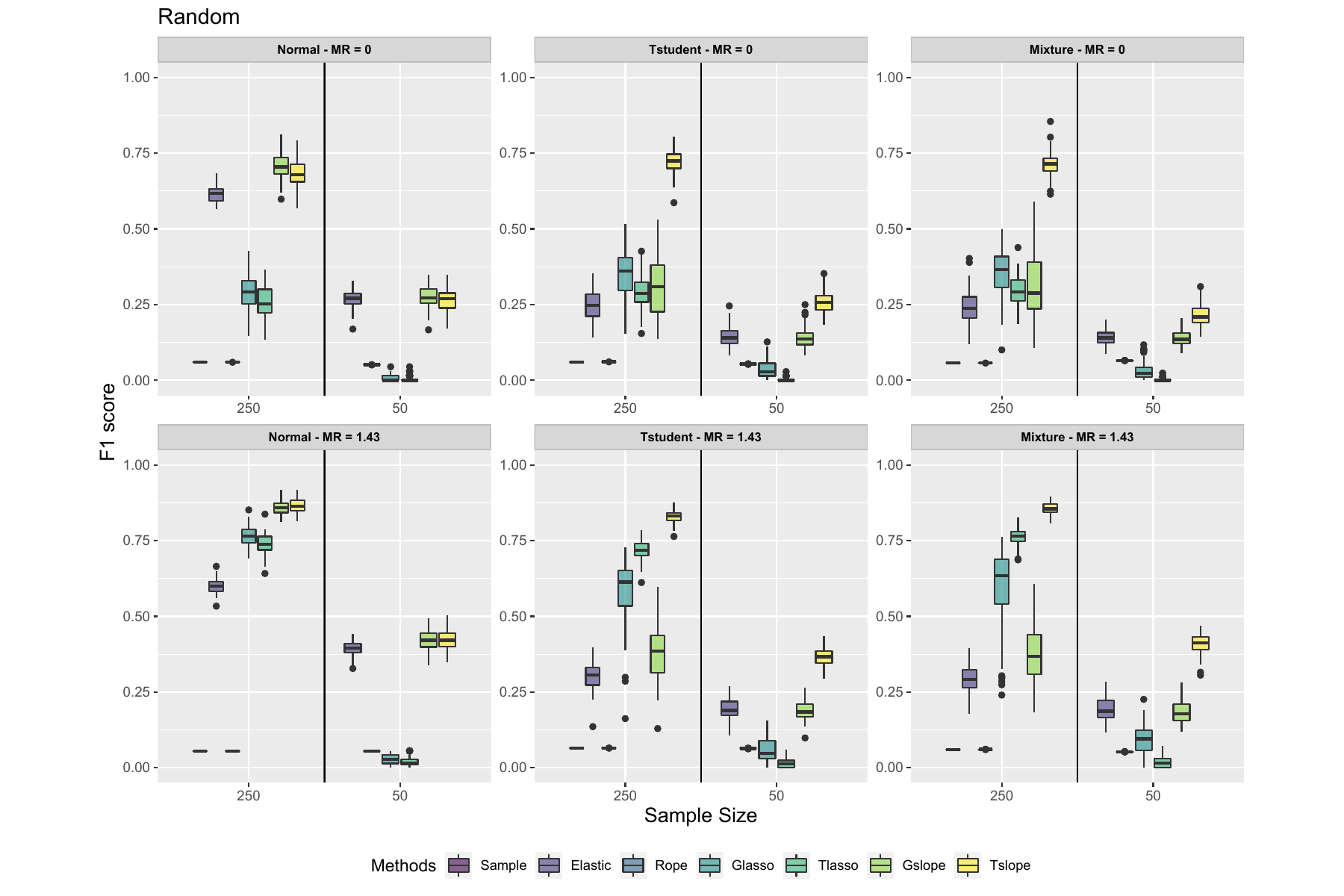}
        \end{tabular}
    \end{adjustbox}
    \caption{F1- Score Boxplots. The figure shows the F-Measure Boxplots across all 100 simulation runs for the 24 distinct network configurations for both the low dimensional ($n>p$) and the high dimensional setting ($n<p$), respectively.}
    \label{fig:f1_box}
\end{figure}
%%% F1 Score - n>p - End %%%%%
%

%
%%% Frobenius Norm - Start %%%%%
\begin{figure}
    \centering
    \begin{adjustbox}{width=\textwidth}
    \begin{tabular}{c}
    \includegraphics[scale=1]{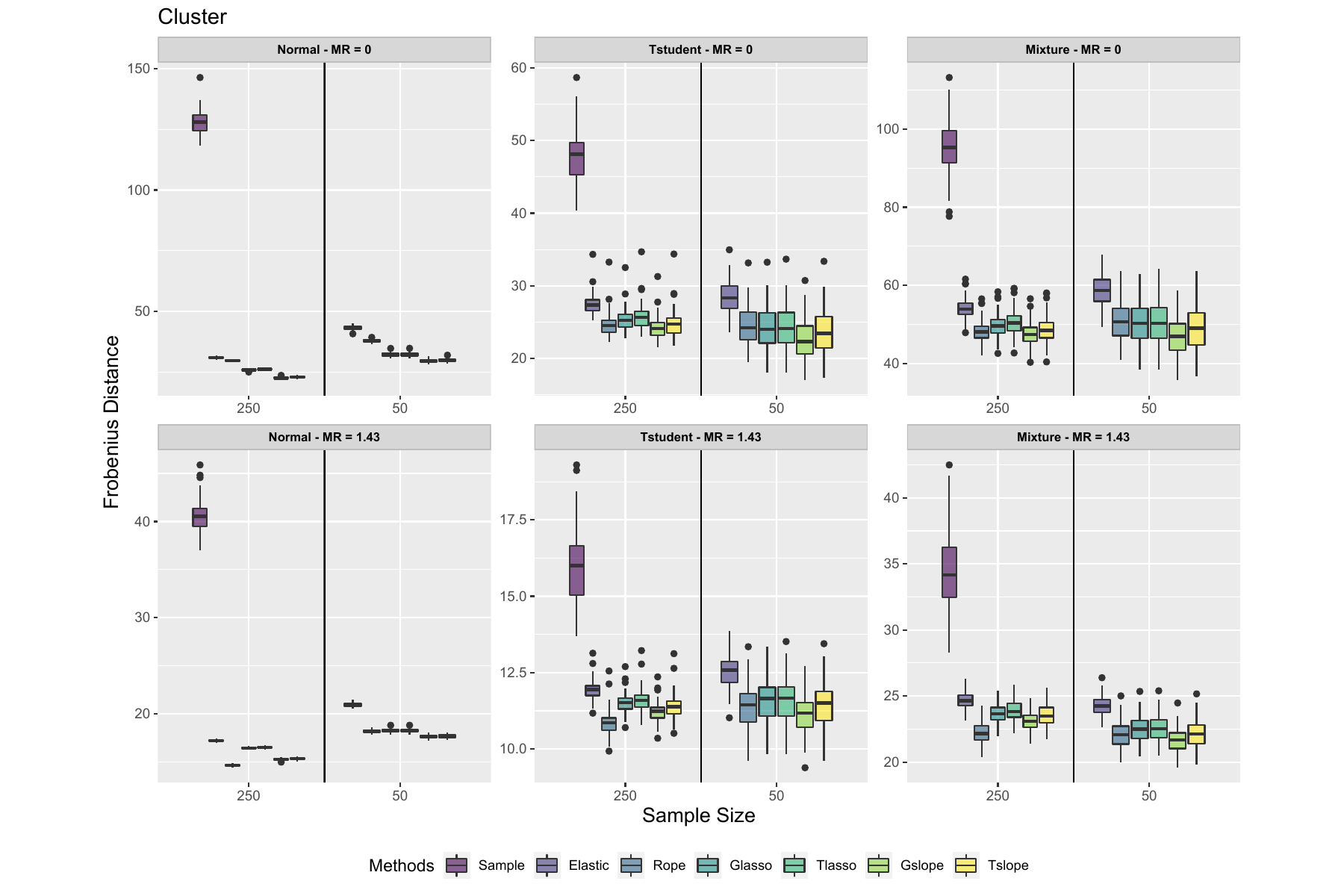} \\
    \includegraphics[scale=1]{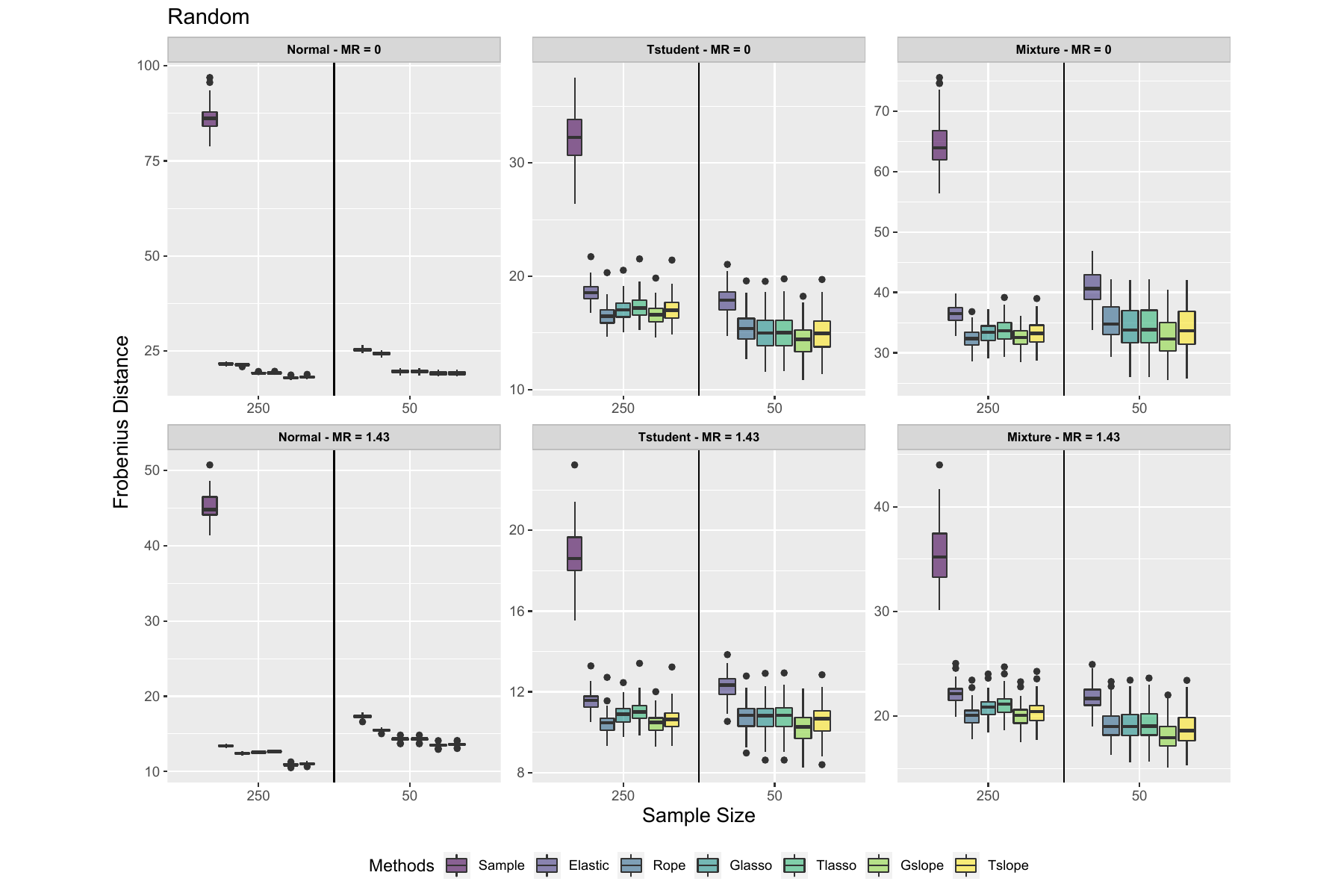}
        \end{tabular}
    \end{adjustbox}
    \caption{Frobenius Norm Distance Boxplots. The figure shows the Frobenius Norm Boxplots across all 100 simulation runs for the 24 distinct network configurations for both the low dimensional ($n>p$) and the high dimensional setting ($n<p$), respectively.}
    \label{fig:frob_box}
\end{figure}
%%% Frobenius Norm - End %%%%%
%

%%% Distant FDR Control - End %%%%%
%

Figure \ref{fig:f1_box} reports the boxplots of the F1-Score across 100 simulations for both the \textit{cluster} (top two rows) and the \textit{random} (bottom two rows) network structure, with MR=0 (1st and 3rd row) and MR=1.43 (2nd and 4th rows), respectively, and considering in each subplot on the left the low dimensional setting with $n>p$ (i.e. $n=250$ and $p=100$), and on the right the high dimensional setting with $n<p$ (i.e. $n=50$ and $p=100$). Boxplots in the first column consider the Gaussian case, in the second column the t-Student and in the 3rd column the Mixture distribution. In each subplot, from left to right, we report the results for the Sample, the Elastic Net, the Rope, the Glasso,  the Tlasso, as well as the Gslope and Tslope procedures.\\
Looking at the results for the low dimensional settings (i.e. $n=250$, $p=100$), both Gslope and Tslope consistently outperform all other methods in the Gaussian case, irrespective of the network configuration and the magnitude ratio. Furthermore, even when considering non-Gaussian distributions, the graphical slope methods perform among the best. Especially, our newly introduced method Tslope shows its adaptivity with respect to the underlying distribution and the presence of fatter tails. While Tslope performs in line with Gslope under the Gaussian setting, it outperforms  Gslope for a t-Student and a mixture distribution. This observation is robust to the network configuration, and towards the assumed magnitude ratio.\\
It is interesting to observe that for this low dimensional setup, both the sample and the Rope methods performs worse than other methods. As both do not impose any sparsity onto the estimated covariance structure, they fail to dissect the underlying graph structure.\\
Reducing the number of observations, while keeping the number of parameters constant ($n=50<p=100$), Figure \ref{fig:f1_box} shows that all methods suffer in extracting the true underlying covariance structure  as compared to the low dimensional case. This is especially true for the Lasso methods, while Elastic Net, Gslope and Tslope perform among the best. In fact, for the cluster network and independent of the magnitude ratio, Gslope and Tslope perform in line with the Elastic Net under the Gaussian assumption, while again Tslope outperforms all methods for non-Gaussian data. These observations also hold for the random network structure.\\
Finally, it can be observed that the performance of the SLOPE procedures (i.e. Gslope and Tslope) is best when the network structure is characterized by truly distinct groups of features. This is evident from the performance of SLOPE methods in the \textit{cluster} network structure with a low magnitude ratio and the \textit{random} network structure with a high magnitude ratio. Reconsidering the heatmaps of Figure \ref{fig:network_structures}, we can see that Panel (a) and Panel (d) will form more distinct groups of features, while the networks in Panel (b) and (c) represent a more in-distinctive setting. As the unit ball of the dual SLOPE norm takes a form of a permutahedron, SLOPE has a natural ability to cluster precision (i.e., also partial correlations) parameters into the groups with the same or very similar value (see e.g., \cite{Schneider2020, Skalski2022, Pattern}). This feature helps the method to better extract the underlying covariance structure in the environments of Panel (a) and (d).\\

%
%%% FDR Control - Start %%%%%

%%% FDR Control - End %%%%%
%

\noindent
Concerning the Frobenius distance, Figure \ref{fig:frob_box} confirms our findings from above, showing that Gslope and Tslope represent the best performing methods across all state-of-the-art sparse graphical modelling approaches. There is only one exception  for a Gaussian distribution in a cluster network and when MR=1.43. Here, the Rope method performs exceptionally well, but still in line with the Slope procedures. \\
%A possible explanation is that this set-up favours the performance of the L2-Norm, while the L1 and sorted L1 norms induce too much sparsity, thereby leading to a higher Frobenius Norm Distance.\\
Comparing among the Gslope and Tslope procedures, Gslope has a marginally better ability to extract the magnitude of the respective edges with a lower variability than Tslope, whereas this observation holds across all of the network configurations and when considering both the low and high dimensional setting. 
While for Gaussian data, as expected, Gslope performs remarkably well, for non-Gaussian data, Tslope stands out, with a better performance in uncovering the relationships among the parameters and only a marginally neglectable inferior performance to Gslope of estimating the magnitude of those relationships.\\
%Summing up, our simulation analysis establishes Gslope and Tslope as two new dominating method among state-of-the art sparse graphical modelling approaches, for both Gaussian and non-Gaussian data thereby controlling the FDR.

\section{Empirical analysis}\label{sec:results}
\subsection{Gene expression data}
\cite{Wille2004} and \cite{Kovacs2021} use graphical models to analyze a real-world dataset made of $p = 39$ expression levels of isoprenoid genes from $n = 118$ samples from the plant Arabidopsis Thaliana. Figure \ref{fig:Gslope_Tslope_Networks} illustrates the results of our analysis of this data set with 
different graphical methods tested in our simulation study, which show that  %These results 
%to identify relevant connections among variables and point out that
differences observed in simulations persist also in these real-world data. 
Firstly, we can observe that Rope and Elastic Net recover very dense graphs, whose structure is rather difficult to analyze.
On the other hand Glasso and Tlasso using $\lambda$ selected according to \cite{Banerjee2008} at the FWER level $\alpha=0.05$ provides graphs which seem to be too sparse. Specifically, they leave out several unconnected nodes, which seem to be rather unrealistic in the gene expression data. The Slope methods are placed in the middle, with
 Gslope identifying a denser graphical model than Tslope. Both methods seem capable of clustering genes that appear to be related (e.g. PPDS1 and PPDS2 or GGPPS2, GGPPS4, GGPPS5, GGPPS8, GGPPS9, GGPPS10) as suggested by \cite{Wille2004}. Following the interpretation by \cite{Wille2004} and results from our simulations, it seems that Tslope can better deal with noise in the data and avoids detecting a larger number of false discoveries, as some of the connections identified by Gslope appear hard to be interpreted from a biological perspective. However, by studying the common edge sets and the difference among the estimated graphical models, we can gain a better insights on which relationship might be persistent and which ones might need further investigation as they are always not detected.\\

%
%%% Empirical Networks - Start %%%%%
%\begin{figure}[h!]
   % \centering
%    \begin{adjustbox}{width=\textwidth}
%    \begin{tabular}{ccc}
%    \includegraphics[scale=1]{Figures/Gslope_new_2.pdf} &
%    \includegraphics[scale=1]{Figures/Tslope_new_2.pdf} &
%    \includegraphics[scale=1]{Figures/Differences_Gslope_Tslope.pdf}\\
%    \end{tabular}
%    \end{adjustbox}
%    \caption{Gene expression network. The Figure shows from left to right the gene expression network of $p=39$ isoprenoid genes using $n=118$ samples from the Arabidopsis Thaliana plant, for the GSlope method, the Tslope method, and for the difference between the GSlope and the Tslope method, respectively.}
    \label{fig:Gslope_Tslope_Networks}
%\end{figure}
%%% Empirical Networks - End %%%%%
%

\begin{figure}[h!]
    \centering
    \begin{adjustbox}{width=\textwidth}
    \begin{tabular}{cc}
    \includegraphics[scale=1]{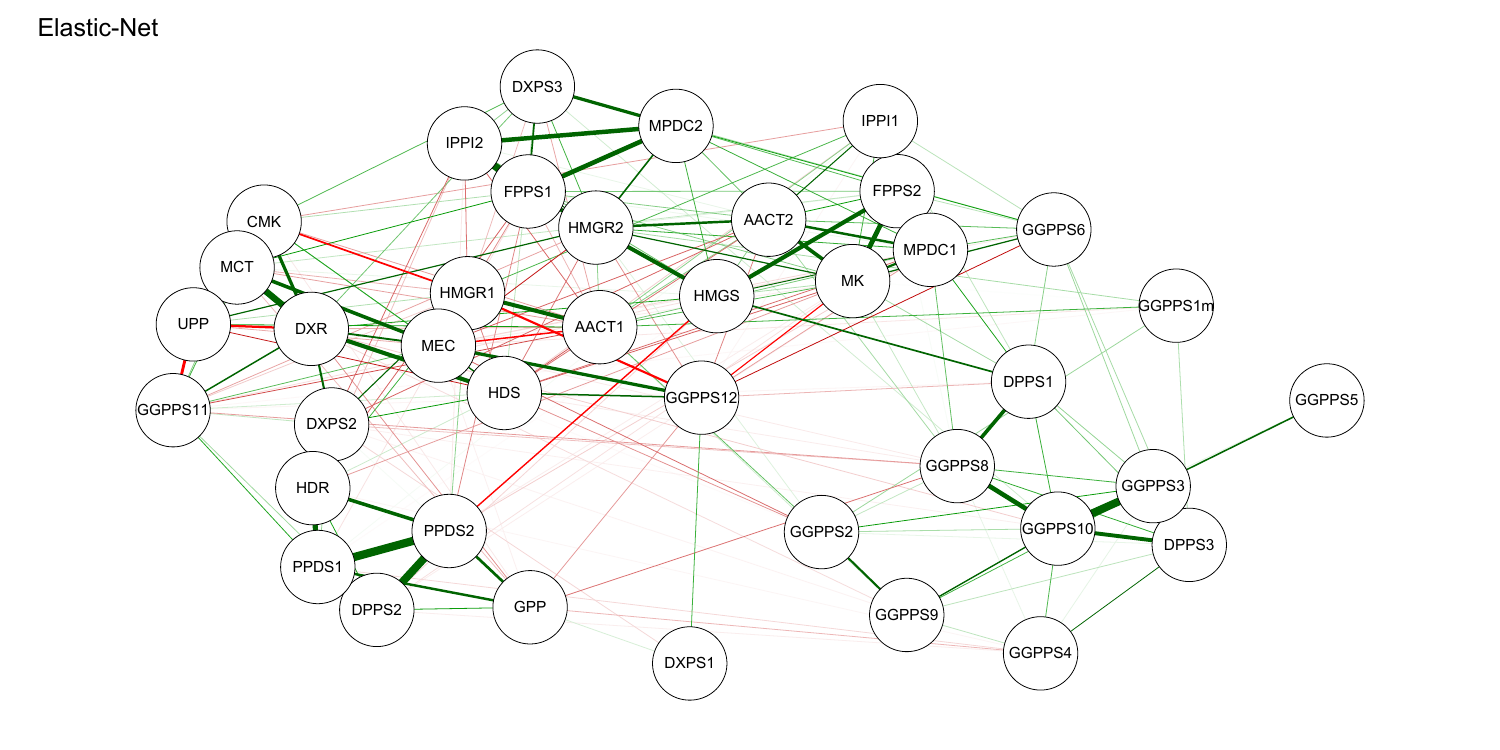} &
    \includegraphics[scale=1]{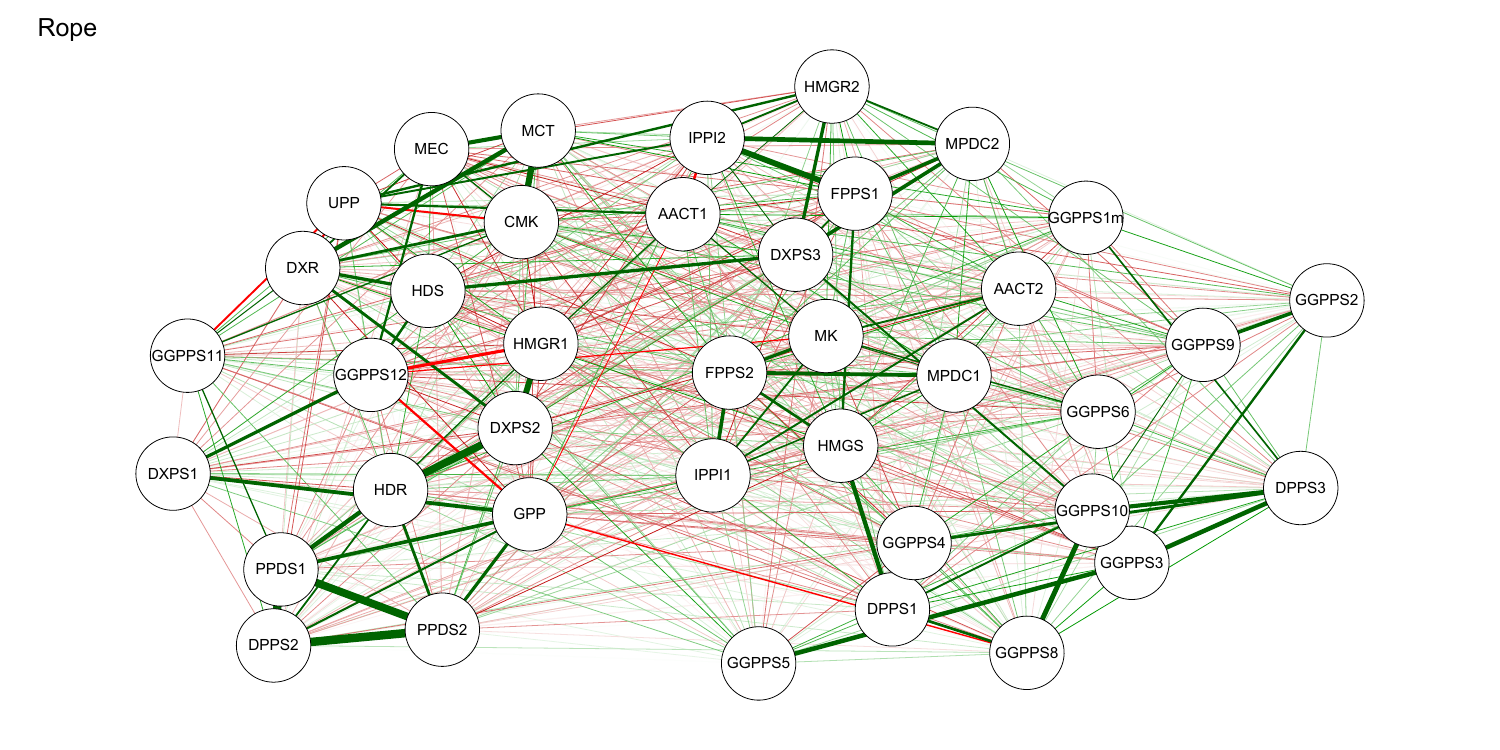} \\
    \includegraphics[scale=1]{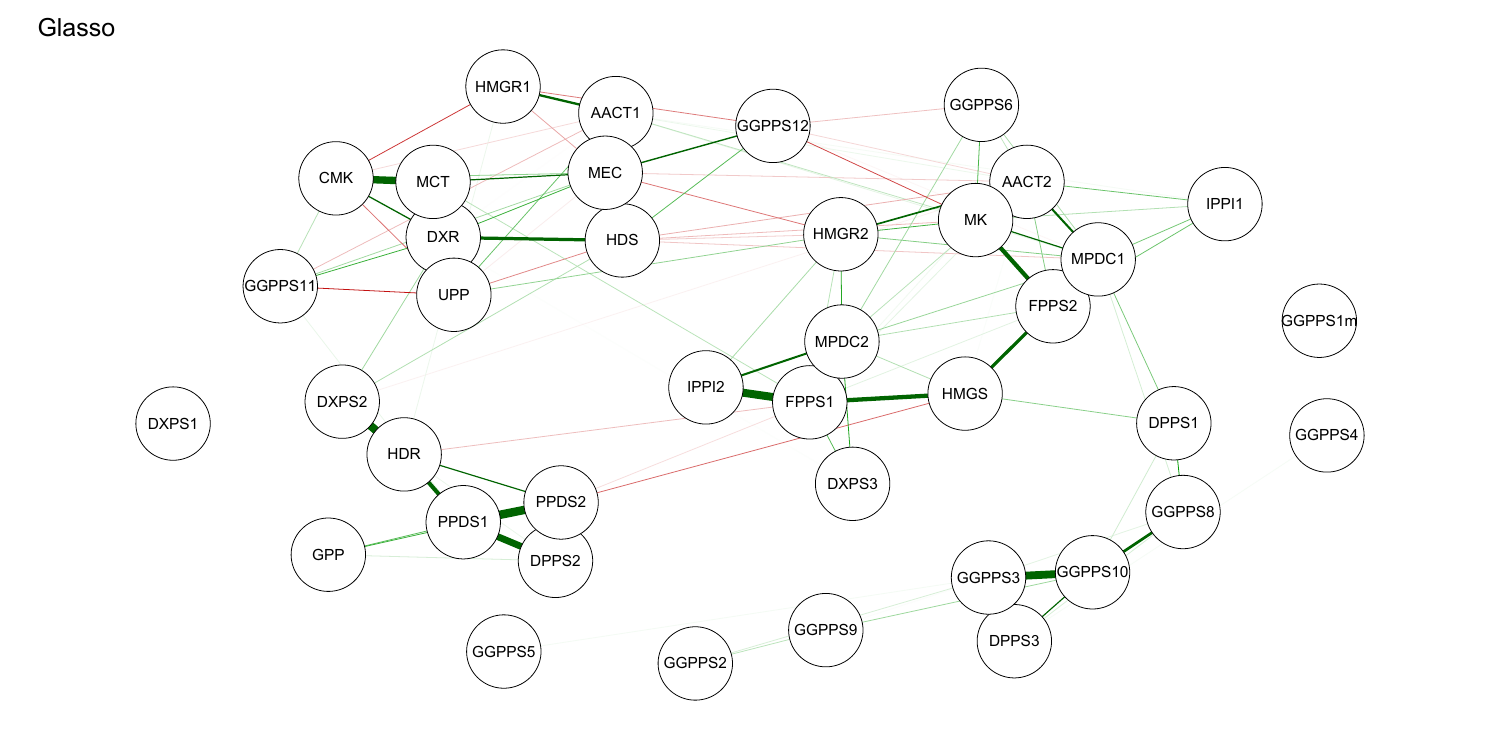}&
    \includegraphics[scale=1]{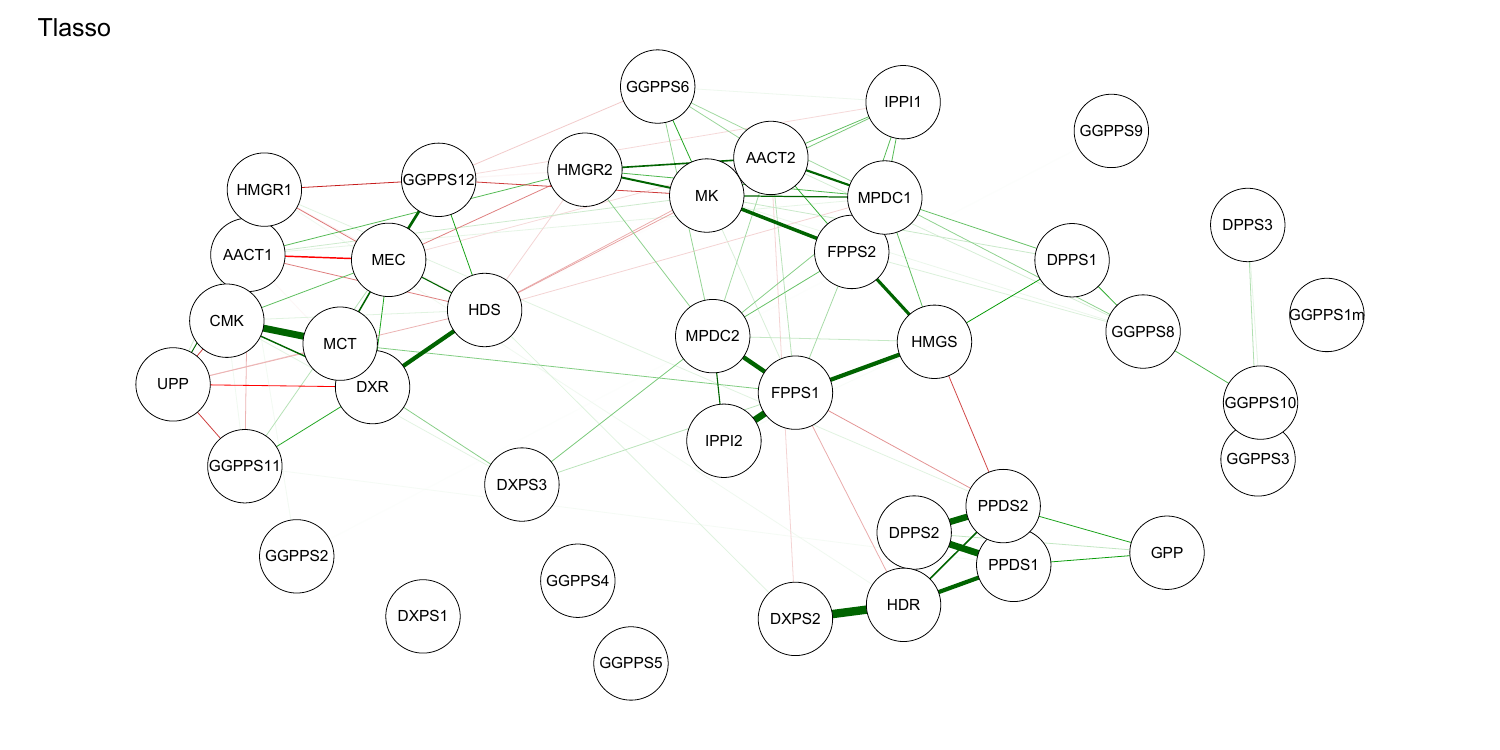}\\
    \includegraphics[scale=1]{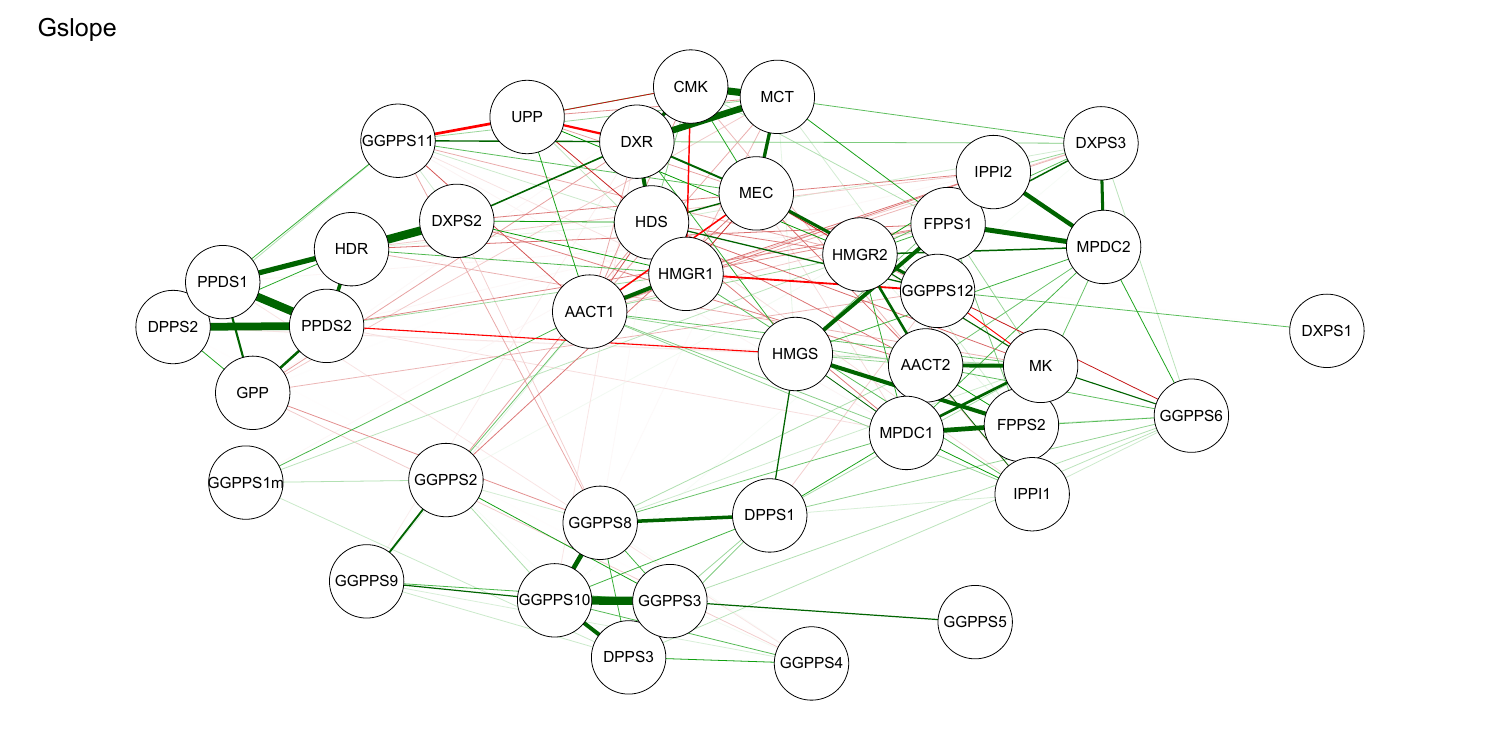} &
    \includegraphics[scale=1]{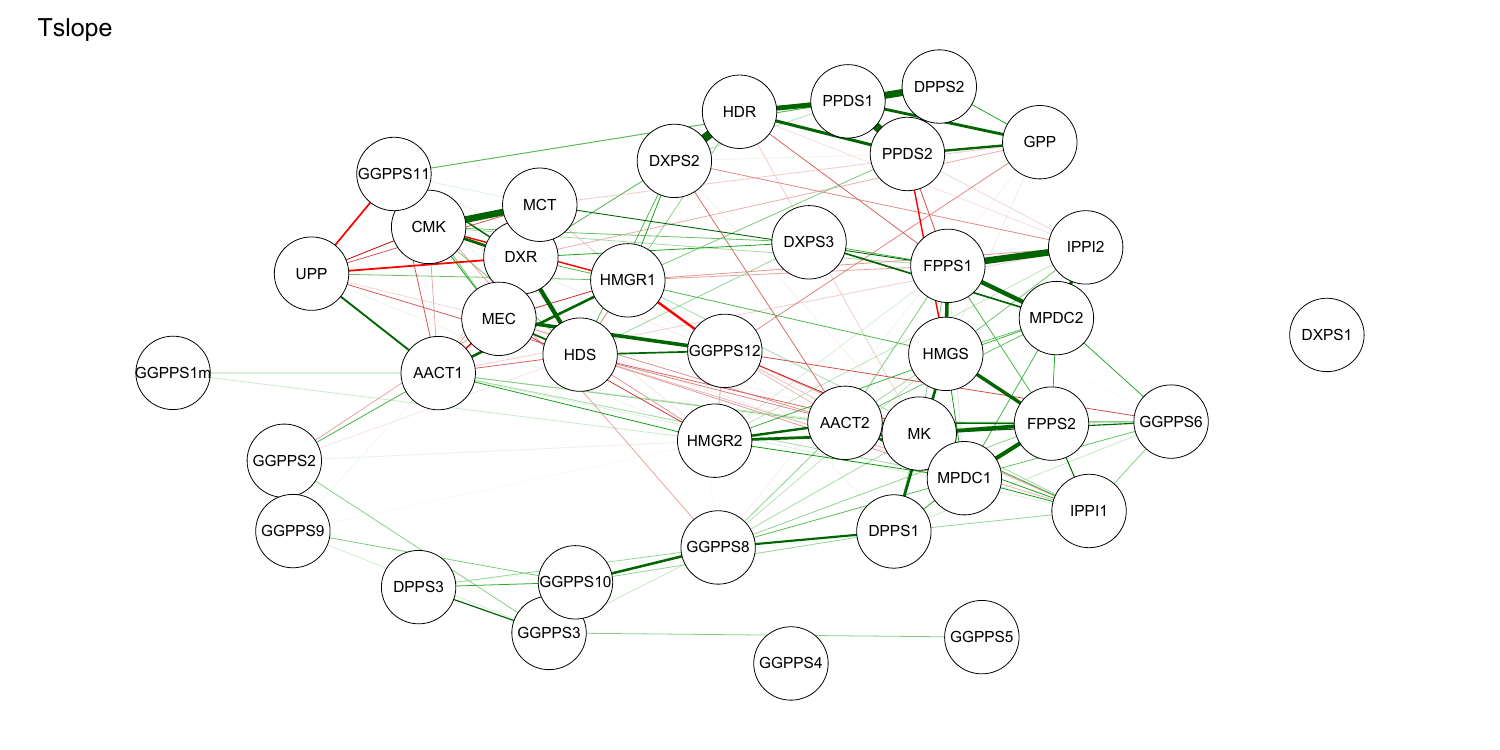}
    \end{tabular}
    \end{adjustbox}
    \caption{Gene expression networks produced by different methods of graphical model selection.}
    \label{fig:Gslope_Tslope_Networks}
\end{figure}

\subsection{Portfolio optimization}

In the parallel article \cite{Riccobello_portfolio} Gslope and Tslope were applied for estimating the assets' precision matrix in the context of portfolio selection.  Extensive simulation and real-world analyses highlight the superiority of our new methods over other state-of-the-art approaches, especially with regard to clustering similar assets and stability characteristics in a high-dimensional scenarios.  The empirical real data analysis further highlights the improvements in terms of risk and risk-adjusted returns provided by Gslope and Tslope for large portfolios, including assets with heavy-tailed distributions. At the same time, these methods show low turnover rates, bringing further advantages with regard to the impact of transaction costs.

\section{Discussion}

In this article, we introduced two novel regularization methods for the estimation of high-dimensional precision matrices, with the penalty defined by the Sorted L-One norm (SLOPE, \cite{Bogdan2013, Bogdan2015}). First of these methods, Gslope, is designed for estimating gaussian graphical models. The second method, Tslope, extends Gslope to the situation when the distribution of the considered random vector has heavy tails. We proposed and implemented an ADMM algorithm for solving the Gslope optimization problem and an EM algorithm for solving Tslope. We also proposed tuning parameters and tuning sequences for Glasso, Gslope and Tslope, with the goal of eliminating false edges connecting different graph connectivity components. Firstly, we relaxed the tuning parameter for Glasso introduced in \cite{Banerjee2008}, so the probability of including such false 'distant' edges is still controlled at assumed level while the power to identify the true edges slightly increases. In case of Gslope, we further relax the strength of regularization by using the slowly decaying sequence of tuning parameters based on the  \cite{Holm_1979} and \cite{Hochberg_1988} multiple testing procedures. This version of Gslope provably controls the probability of including the false edges between different connectivity components under the standard assumptions of the validity of the Hochberg procedure. We also introduce a quickly decaying sequence based on the \cite{Benjamini1995} multiple testing procedure, which has been empirically shown to control the percentage of false 'distant' edges  among all discovered edges (distant FDR control), both for Gslope and Tslope , and yields much higher power of identifying important edges.
Our empirical studies illustrate that the respective versions of Gslope and Tslope outperform other state-of-the-art methods with respect to the accuracy of identifying the graph structure and the precision of the estimation of the concentration matrix, with Tslope outperforming Gslope when the distribution of the considered random vector has heavy tails. Good properties of our methods have been confirmed by the analysis of real biological and financial data.

Concerning the future development, it would be of interest to further investigate theoretical properties of Gslope and Tslope. Specifically, the results included in \cite{Kos2019, Kos2020} on the False Discovery Rate (FDR) control of the generalized versions of SLOPE suggest the direction for the formal proof of the asymptotic 'distant' FDR control by Gslope. Moreover, it would be interesting to extend novel results from \cite{Pattern} on the pattern recovery by SLOPE into the context of graphical models. These new results provide the conditions under which SLOPE can properly identify the low dimensional model by eliminating parameters which are equal to zero and by equalizing estimators of parameters, which are equal to each other. As reported in \cite{Pattern}, this additional level of dimensionality reduction allows for a substantial improvement of the estimation accuracy with respect to LASSO. The practical advantages of the clustering properties of Gslope and Tslope in the context of portfolio optimization have been reported in \cite{Riccobello_portfolio}.

While the theoretical results on the FDR control and the pattern recovery by SLOPE are very encouraging, it must be noted that they hold under rather stringent assumptions. For example, Gslope can control the number of false edges between the distinct connectivity components but it would be rather difficult (or impossible) to construct the sequence of the tuning parameters to control FDR within the connectivity components. Also, the irrepresentability condition for the pattern recovery by SLOPE is rather restrictive (see \cite{Pattern}). However, as discussed in \cite{adaptiveSL,Pattern,Geom2}, the model selection properties of SLOPE can be substantially improved by using its adaptive version or by clustering values of similar SLOPE estimators (thresholded version). Specifically, theoretical and empirical results from \cite{adaptiveSL,Pattern, Geom2} suggest that these versions of SLOPE can recover the true model under much weaker assumptions than adaptive or thresholded LASSO. Moreover, adaptive Bayesian version of SLOPE \citep{adaptiveSL} allows for increased efficiency by incorporating the prior knowledge on the model structure. In the future it would be of interest to develop appropriate adaptive or thresholded versions of Gslope and Tslope and investigate their properties.

%\subsection{Financial data application}

%In the parallel article \cite{} we applied 

%\newpage
%\section{Conclusion, discussion and future research}\label{sec:conclusion}

% Acknowledgements should go at the end, before appendices and references
%\acks{
%%First author would like to thank his thesis advisor Sandra for guidance and the opportunity to fulfill a small dream of writing this research paper. 
%MB acknowledges the  support of the Swedish Research Council, grant no.
%2020-05081 and thanks Michał Makowski for extensive simulations of Gslope during in his Master Thesis research. 
%%Thanks to her guide and supported by my co-authors I have had the occasion to fulfilling a small dream.

\newpage

%\bigskip
%\begin{center}
%{\large\bf SUPPLEMENTARY MATERIAL}
%\end{center}
%
%\begin{description}
%
%\item[Title:] Brief description. (file type)
%
%\item[R-package for  MYNEW routine:] R-package ÒMYNEWÓ containing code to perform the diagnostic methods described in the article. The package also contains all datasets used as examples in the article. (GNU zipped tar file)
%
%\item[HIV data set:] Data set used in the illustration of MYNEW method in Section~ 3.2. (.txt file)
%
%\end{description}
%
%\section{BibTeX}
%
%We hope you've chosen to use BibTeX!\ If you have, please feel free to use the package natbib with any bibliography style you're comfortable with. The .bst file Chicago was used here, and agsm.bst has been included here for your convenience. 

\bibliographystyle{plainnat}
\bibliography{bibfile.bib}

\newpage

\appendix

\bigskip
\begin{center}
{\large\bf APPENDIX}
\end{center}

\section{Alternating Direction Method of Multipliers}\label{sec:ADMM}

The Alternating Direction Method of Multipliers (ADMM) represents an effective tool to solve convex optimization problems \citep{Boyd2011}. 
%It combines  the Dual Ascent and the Augmented Lagrangian methods. 
%%%%%%%%%%%%%%%%%%%%%%%%%%%%%%%%%%%%%%%%%%%%%%%%%%%%%%%%%%%%%%%%%%%%%%%%%%%%%%%%%%%%%%%%%%%
%%%%%%%%%%%%%%%%%%%%%%%%%%%%%%%%%%%%%%%%%%%%%%%%%%%%%%%%%%%%%%%%%%%%%%%%%%%%%%%%%%%%%%%%%%%
%%%%%%%%%%%%%%%%%%%%%%%%%%%%%%%%%%%%%%%%%%%%%%%%%%%%%%%%%%%%%%%%%%%%%%%%%%%%%%%%%%%%%%%%%%%
%%%%%%%%%%%%%%%%%%%%%%%%%%%Algorithm description%%%%%%%%%%%%%%%%%%%%%%%%%%%%%%%%%%%%%%%%%%
%%%%%%%%%%%%%%%%%%%%%%%%%%%%%%%%%%%%%%%%%%%%%%%%%%%%%%%%%%%%%%%%%%%%%%%%%%%%%%%%%%%%%%%%%%%
%%%%%%%%%%%%%%%%%%%%%%%%%%%%%%%%%%%%%%%%%%%%%%%%%%%%%%%%%%%%%%%%%%%%%%%%%%%%%%%%%%%%%%%%%%%
%%%%%%%%%%%%%%%%%%%%%%%%%%%%%%%%%%%%%%%%%%%%%%%%%%%%%%%%%%%%%%%%%%%%%%%%%%%%%%%%%%%%%%%%%%%
The ADMM algorithm has the convergence properties of the method of multipliers. Moreover, it also possesses the decomposability of the dual ascent. It solves problems that can be expressed in the following form:
\begin{equation}
\begin{cases}
\min\quad & f_{0}(\boldsymbol{x})+g_{0}(\boldsymbol{y})\\
s.t.\quad & \boldsymbol{A}\boldsymbol{x}+\boldsymbol{B}\boldsymbol{y}=\boldsymbol{c}
\end{cases},\label{eq: 5.1}
\end{equation}
where $\boldsymbol{x}\in\mathbb{R}_{n\times1}$, $\boldsymbol{y}\in\mathbb{R}_{m\times1}$, $\boldsymbol{A}\in\mathbb{R}_{p\times n}$, $\boldsymbol{B}\in\mathbb{R}_{p\times m}$, $\boldsymbol{c}\in\mathbb{R}_{p\times1}$. 

%Comparing the problem (\ref{eq: 5.1}) with the problem (\ref{eq: A.19}), here the main variable has been split into two parts, $\boldsymbol{x}$ and $\boldsymbol{y}$, the same is valid for the objective function. 
Our aim is to find the optimal pair of values $(\boldsymbol{x}^{*},\boldsymbol{y}^{*})$ which solves the problem in \eqref{eq: 5.1}, given the equality constraints. The corresponding augmented Lagrangian function is defined as:
\begin{equation}
\mathcal{L}^{+}(\boldsymbol{x},\boldsymbol{y},\boldsymbol{\lambda})=f_{0}(\boldsymbol{x})+g_{0}(\boldsymbol{y})+\boldsymbol{\lambda}^\prime(\boldsymbol{A}\boldsymbol{x}+\boldsymbol{B}\boldsymbol{y}-\boldsymbol{c})+\frac{\rho}{2}\left\Vert \boldsymbol{A}\boldsymbol{x}+\boldsymbol{B}\boldsymbol{y}-\boldsymbol{c}\right\Vert_{2}^{2}.
\end{equation}

The ADMM algorithm consists of the following iterations:
\begin{equation}
\begin{cases}
\boldsymbol{x}_{(k+1)}\coloneqq & \arg\min_{\boldsymbol{x}}\mathcal{L^{+}}\left(\boldsymbol{x},\boldsymbol{y}_{(k)},\boldsymbol{\lambda}_{(k)}\right)\\
\boldsymbol{y}_{(k+1)}\coloneqq & \arg\min_{\boldsymbol{y}}\mathcal{L^{+}}\left(\boldsymbol{x}_{(k+1)},\boldsymbol{y},\boldsymbol{\lambda}_{(k)}\right)\\
\boldsymbol{\lambda}_{(k+1)}\coloneqq & \boldsymbol{\lambda}_{(k)}+\rho\left(\boldsymbol{A}\boldsymbol{x}_{(k+1)}+\boldsymbol{B}\boldsymbol{y}_{(k+1)}-\boldsymbol{c}\right)
\end{cases}.
\end{equation}

%The reader who has read appendix (\ref{subsec: Appendix A.4.2}), has surely note that the algorithm is nothing more than the blend of the dual ascent and the method of multipliers: we have the $x$-minimization step, the $y$-minimization step and the dual update. 
Similar to the method of multipliers, the Lagrange multiplier is updated using the step size equal to the augmented Lagrangian penalty parameter $\rho$. The variables $\boldsymbol{x}$ and $\boldsymbol{y}$ are alternatively updated, and this explains the denomination of the ADMM algorithm.

%%%%%%%%%%%%%%%%%%%%%%%%%%%%%%%%%%%%%%%%%%%%%%%%%%%%%%%%%%%%%%%%%%%%%%%%%%%%%%%%%%%%%%%%%%%
%%%%%%%%%%%%%%%%%%%%%%%%%%%%%%%%%%%%%%%%%%%%%%%%%%%%%%%%%%%%%%%%%%%%%%%%%%%%%%%%%%%%%%%%%%%
%%%%%%%%%%%%%%%%%%%%%%%%%%%%%%%%%%%%%%%%%%%%%%%%%%%%%%%%%%%%%%%%%%%%%%%%%%%%%%%%%%%%%%%%%%%
%%%%%%%%%%%%%%%%%%%%%%%%%%%Convergence properties%%%%%%%%%%%%%%%%%%%%%%%%%%%%%%%%%%%%%%%%%%
%%%%%%%%%%%%%%%%%%%%%%%%%%%%%%%%%%%%%%%%%%%%%%%%%%%%%%%%%%%%%%%%%%%%%%%%%%%%%%%%%%%%%%%%%%%
%%%%%%%%%%%%%%%%%%%%%%%%%%%%%%%%%%%%%%%%%%%%%%%%%%%%%%%%%%%%%%%%%%%%%%%%%%%%%%%%%%%%%%%%%%%
%%%%%%%%%%%%%%%%%%%%%%%%%%%%%%%%%%%%%%%%%%%%%%%%%%%%%%%%%%%%%%%%%%%%%%%%%%%%%%%%%%%%%%%%%%%

It is important to highlight the fact that,   
%We will present only the most important convergence results, the reader interested in proof and formal results can refer to \citep{36}. 
under the two assumptions: i) functions $f_{0}(\boldsymbol{x})$ and $g_{0}(\boldsymbol{y})$ are closed, proper and convex; and ii) the augmented Lagrangian function $\mathcal{L}^{+}$ with penalty $\rho=0$ has a saddle point, the ADMM iterates satisfy the following properties:
\begin{enumerate}
\item \textit{Residual convergence}: the residual, defined as $(\boldsymbol{A}\boldsymbol{x}+\boldsymbol{B}\boldsymbol{y}-\boldsymbol{c})_{(k)}$,
approaches zero as $k\rightarrow\infty$;
\item \textit{Objective convergence}: $f_{0}(\boldsymbol{x}_{(k)})+g_{0}(\boldsymbol{y}_{(k)})\rightarrow f_{0}(\boldsymbol{x}^{*})+g_{0}(\boldsymbol{y}^{*})$
as $k\rightarrow\infty$. In other words the objective function approaches its optimal value;
\item \textit{Dual variable convergence}: $\boldsymbol{\lambda}_{(k)}\rightarrow\boldsymbol{\lambda}^{*}$
as $k\rightarrow\infty$, where $\boldsymbol{\lambda}^{*}$ is a dual
optimal point.
\end{enumerate}
%In practice, simple experiments show that ADMM can be very slow to
%converge to high accuracy. However, it is often the case that ADMM
%converges to modest accuracy, within a few tens of iterations. This
%is not a problem for our purpose, in statistical and machine learning
%problems, solving a parameter estimation problem to very high accuracy
%often yields little to no improvement in actual prediction performance

%%%%%%%%%%%%%%%%%%%%%%%%%%%%%%%%%%%%%%%%%%%%%%%%%%%%%%%%%%%%%%%%%%%%%%%%%%%%%%%%%%%%%%%%%%%
%%%%%%%%%%%%%%%%%%%%%%%%%%%%%%%%%%%%%%%%%%%%%%%%%%%%%%%%%%%%%%%%%%%%%%%%%%%%%%%%%%%%%%%%%%%
%%%%%%%%%%%%%%%%%%%%%%%%%%%%%%%%%%%%%%%%%%%%%%%%%%%%%%%%%%%%%%%%%%%%%%%%%%%%%%%%%%%%%%%%%%%
%%%%%%%%%%%%%%%%%%%%%%%%%%%Optimality conditions%%%%%%%%%%%%%%%%%%%%%%%%%%%%%%%%%%%%%%%%%%
%%%%%%%%%%%%%%%%%%%%%%%%%%%%%%%%%%%%%%%%%%%%%%%%%%%%%%%%%%%%%%%%%%%%%%%%%%%%%%%%%%%%%%%%%%%
%%%%%%%%%%%%%%%%%%%%%%%%%%%%%%%%%%%%%%%%%%%%%%%%%%%%%%%%%%%%%%%%%%%%%%%%%%%%%%%%%%%%%%%%%%%
%%%%%%%%%%%%%%%%%%%%%%%%%%%%%%%%%%%%%%%%%%%%%%%%%%%%%%%%%%%%%%%%%%%%%%%%%%%%%%%%%%%%%%%%%%%

The following necessary and sufficient conditions are required to solve the problem in (\ref{eq: 5.1}):
%In order to solve problem (\ref{eq: 5.1}) the following necessary and sufficient optimality conditions need to hold:
\begin{equation}
\boldsymbol{A}\boldsymbol{x}^{*}+\boldsymbol{B}\boldsymbol{y}^{*}-\boldsymbol{c}=\boldsymbol{0}\label{eq: 5.4},
\end{equation}
\begin{equation}
\begin{cases}
\boldsymbol{0}\in & \partial f(\boldsymbol{x}^{*})+\boldsymbol{A}^\prime\boldsymbol{\lambda}^{*}\\
\boldsymbol{0}\in & \partial f(\boldsymbol{y}^{*})+\boldsymbol{B}^\prime\boldsymbol{\lambda}^{*}
\end{cases},\label{eq: 5.5}
\end{equation}
where $\partial$ denotes the sub-differential operator.\footnote{Note that when functions $f$ and $g$ are differentiable, the sub-differentials can be replaced by the gradients $\nabla f$ and $\nabla g$. Likewise, the symbol `$\in$' can be replaced by `$=$'. As a result, the set of equations in \eqref{eq: 5.5} can be rewritten as follows:
\[
\begin{cases}
\boldsymbol{0}= & \nabla f(\boldsymbol{x}^{*})+\boldsymbol{A}^\prime\boldsymbol{\lambda}^{*}\\
\boldsymbol{0}= & \nabla f(\boldsymbol{y}^{*})+\boldsymbol{B}^\prime\boldsymbol{\lambda}^{*}
\end{cases}.
\]
}

Equation \eqref{eq: 5.4} is the primal feasibility condition. In contrast, the set of relationships in \eqref{eq: 5.5} represents the dual feasibility condition. The augmented Lagrangian function $\mathcal{L^{+}}\left(\boldsymbol{x}_{(k+1)},\boldsymbol{y},\boldsymbol{\lambda}_{(k)}\right)$ is minimized by $\boldsymbol{y}_{(k+1)}$, which leads to the following result:

\begin{eqnarray}
\boldsymbol{0} &\in & \partial f\left(\boldsymbol{y}_{(k+1)}\right)+\boldsymbol{B}^\prime\boldsymbol{\lambda}_{(k)}+\rho\boldsymbol{B}^\prime \left(\boldsymbol{A}\boldsymbol{x}_{(k+1)}+\boldsymbol{B}\boldsymbol{y}_{(k+1)}-\boldsymbol{c}\right) \nonumber \\ 
&=& \partial f\left(\boldsymbol{y}_{(k+1)}\right)+\boldsymbol{B}^\prime\left[\boldsymbol{\lambda}_{(k)}+\rho\left(\boldsymbol{A}\boldsymbol{x}_{(k+1)}+\boldsymbol{B}\boldsymbol{y}_{(k+1)}-\boldsymbol{c}\right)\right] \nonumber \\
&=&  \partial f\left(\boldsymbol{y}_{(k+1)}\right)+\boldsymbol{B}^\prime\boldsymbol{\lambda}_{(k+1)}. 
\end{eqnarray}
%\begin{align}
%\boldsymbol{0}\in\: & \partial f\left(\boldsymbol{y}^{k+1}\right)+\boldsymbol{B}^\prime\boldsymbol{\lambda}^{k}+\rho\boldsymbol{B}^\prime \left(\boldsymbol{A}\boldsymbol{x}^{k+1}+\boldsymbol{B}\boldsymbol{y}^{k+1}-\boldsymbol{c}\right)\nonumber \\
%=\: & \partial f(\boldsymbol{y}^{k+1})+\boldsymbol{B}^{T}[\boldsymbol{\lambda}^{k}+\rho(\boldsymbol{A}\boldsymbol{x}^{k+1}+\boldsymbol{B}\boldsymbol{y}^{k+1}-\boldsymbol{c})]\\
%=\: & \partial f(\boldsymbol{y}^{k+1})+\boldsymbol{B}^{T}\boldsymbol{\lambda}^{k+1}\nonumber 
%\end{align}

This means that, at each iteration, $\boldsymbol{y}_{(k+1)}$ and $\boldsymbol{\lambda}_{(k+1)}$
satisfy the dual feasibility condition $\boldsymbol{0}\in\partial f(\boldsymbol{y}^{*})+\boldsymbol{B}^\prime\boldsymbol{\lambda}^{*}$. We need to satisfy the remaining conditions to achieve the optimal solution. Given that $\boldsymbol{x}_{(k+1)}$ minimizes the augmented Lagrangian function $\mathcal{L^{+}}\left(\boldsymbol{x},\boldsymbol{y}_{(k)},\boldsymbol{\lambda}_{(k)}\right)$, by definition, we obtain:
\vspace{-0.2cm}
\begin{eqnarray}\label{eq: 5.7}
\boldsymbol{0} & \in & \partial f\left(\boldsymbol{x}_{(k+1)}\right)+\boldsymbol{A}^\prime\boldsymbol{\lambda}_{(k)}+\rho\boldsymbol{A}^\prime\left(\boldsymbol{A}\boldsymbol{x}_{(k+1)}+\boldsymbol{B}\boldsymbol{y}_{(k)}-\boldsymbol{c}\right) \nonumber \\ 
&=& \partial f\left(\boldsymbol{x}_{(k+1)}\right)+\boldsymbol{A}^\prime\left[\boldsymbol{\lambda}_{(k)}+\rho\left(\boldsymbol{A}\boldsymbol{x}_{(k+1)}+\boldsymbol{B}\boldsymbol{y}_{(k+1)}-\boldsymbol{c}\right)+\rho\boldsymbol{B}\left(\boldsymbol{y}_{(k)}-\boldsymbol{y}_{(k+1)}\right)\right] \nonumber \\ 
&=& \partial f\left(\boldsymbol{x}_{(k+1)}\right)+\boldsymbol{A}^\prime\boldsymbol{\lambda}_{(k+1)}+\rho\boldsymbol{A}^\prime\boldsymbol{B}\left(\boldsymbol{y}_{(k)}-\boldsymbol{y}_{(k+1)}\right) .
\end{eqnarray}

We can rewrite the relationship in \eqref{eq: 5.7} as: $\rho\boldsymbol{A}^\prime\boldsymbol{B}\left(\boldsymbol{y}_{(k+1)}-\boldsymbol{y}_{(k)}\right)\in\partial f\left(\boldsymbol{x}_{(k+1)}\right)+\boldsymbol{A}^\prime\boldsymbol{\lambda}_{(k+1)}$
%\[
%\rho\boldsymbol{A}^{T}\boldsymbol{B}(\boldsymbol{y}^{k+1}-\boldsymbol{y}^{k})\in\partial f(\boldsymbol{x}^{k+1})+\boldsymbol{A}^{T}\boldsymbol{\lambda}^{k+1}
%\]
and
define its left side as: $\boldsymbol{s}_{(k+1)}=\rho\boldsymbol{A}^\prime\boldsymbol{B}\left(\boldsymbol{y}_{(k+1)}-\boldsymbol{y}_{(k)}\right)$.
%\[
%\boldsymbol{s}^{k+1}=\rho\boldsymbol{A}^{T}\boldsymbol{B}(\boldsymbol{y}^{k+1}-\boldsymbol{y}^{k})
%\]
This quantity is the dual residual of the dual feasibility condition $\boldsymbol{0}\in\partial f\left(\boldsymbol{x}^{*}\right)+\boldsymbol{A}^\prime\boldsymbol{\lambda}^{*}$ at iteration $k+1$. In contrast, $\boldsymbol{A}\boldsymbol{x}_{(k+1)}+\boldsymbol{B}\boldsymbol{y}_{(k+1)}-\boldsymbol{c}$ is the primal residual at iteration $k+1$. If the primal and dual residuals converge to zero, the optimality conditions are satisfied and the ADMM algorithm converges. 
%(the formal proof is in appendix A of \citet{36}). 
In practical applications, we stop the algorithm when the dual and primal residuals satisfy a given level of tolerance.
%%%%%%%%%%%%%%%%%%%%%%%%%%%%%%%%%%%%%%%%%%%%%%%%%%%%%%%%%%%%%%%%%%%%%%%%%%%%%%%%%%%%%%%%%%%
%%%%%%%%%%%%%%%%%%%%%%%%%%%%%%%%%%%%%%%%%%%%%%%%%%%%%%%%%%%%%%%%%%%%%%%%%%%%%%%%%%%%%%%%%%%
%%%%%%%%%%%%%%%%%%%%%%%%%%%%%%%%%%%%%%%%%%%%%%%%%%%%%%%%%%%%%%%%%%%%%%%%%%%%%%%%%%%%%%%%%%%
%%%%%%%%%%%%%%%%%%%%%%%%%%%ADMM pseudo code%%%%%%%%%%%%%%%%%%%%%%%%%%%%%%%%%%%%%%%%%%
%%%%%%%%%%%%%%%%%%%%%%%%%%%%%%%%%%%%%%%%%%%%%%%%%%%%%%%%%%%%%%%%%%%%%%%%%%%%%%%%%%%%%%%%%%%
%%%%%%%%%%%%%%%%%%%%%%%%%%%%%%%%%%%%%%%%%%%%%%%%%%%%%%%%%%%%%%%%%%%%%%%%%%%%%%%%%%%%%%%%%%%
%%%%%%%%%%%%%%%%%%%%%%%%%%%%%%%%%%%%%%%%%%%%%%%%%%%%%%%%%%%%%%%%%%%%%%%%%%%%%%%%%%%%%%%%%%%
Building on the results presented above, we conclude this section by providing a compact formulation of the ADMM algorithm.  
%We will conclude with the general pseudocode formulation of ADMM:\\

\begin{algorithm}[H]
\SetAlgoLined

$\boldsymbol{y}_{(0)}\leftarrow\boldsymbol{\tilde{y}}$,\:$\boldsymbol{\lambda}_{(0)}\leftarrow\boldsymbol{\tilde{\lambda}}$,\:$\rho\leftarrow\rho_{(0)}>0$,\:$k\leftarrow1$\;
	
	\While{convergence criterion is not satisfied}
		{$\boldsymbol{x}_{(k+1)}\coloneqq\arg\min_{\boldsymbol{x}}\mathcal L^{+}(\boldsymbol{x},\boldsymbol{y}_{(k)},\boldsymbol{\lambda}_{(k)})$\;  
		$\boldsymbol{y}_{(k+1)}\coloneqq\arg\min_{\boldsymbol{y}}\mathcal L^{+}(\boldsymbol{x}_{(k+1)},\boldsymbol{y},\boldsymbol{\lambda}_{(k)})$\; 
		$\boldsymbol{\lambda}_{(k+1)}\coloneqq\boldsymbol{\lambda}_{(k+1)}+\rho(\boldsymbol{Ax}_{(k)}+\boldsymbol{By}_{(k)}-\boldsymbol{c})$\;}

\caption{Alternating Direction Method of Multipliers} \end{algorithm}

\section{Derivation of Equation \eqref{eq:upomega}}\label{sec:omegaupdate}

We report below the derivation of Equation \eqref{eq:upomega}:  
\begin{eqnarray}
\bfTheta_{(k+1)}= & \arg\min_{\bfTheta > \textbf{0}} & \mathcal{L}^{+}\left(\bfTheta,\boldsymbol{Y}_{(k)},\boldsymbol{Z}_{(k)}\right)\nonumber \\
= & \arg\min_{\bfTheta> \textbf{0}} & -\log\det\bfTheta+\tr\left(\bfTheta\boldsymbol{S}\right)+J_{\bflambda}\left(\boldsymbol{Y}_{(k)}\right)+\rho\left\langle \boldsymbol{Z}_{(k)},\bfTheta-\boldsymbol{Y}_{(k)}\right\rangle _{F}\nonumber \\
 &  & +\frac{\rho}{2}\left\Vert\bfTheta-\boldsymbol{Y}_{(k)}\right\Vert_{F}^{2}\nonumber \\
= & \arg\min_{\bfTheta > \textbf{0}} & -\log\det\bfTheta+\left\langle \bfTheta,\boldsymbol{S}\right\rangle _{F}
+\rho\left\langle \boldsymbol{Z}_{(k)},\bfTheta\right\rangle _{F}\nonumber +\frac{\rho}{2}\left\Vert\bfTheta-\boldsymbol{Y}_{(k)}\right\Vert_{F}^{2}\nonumber \\
= & \arg\min_{\bfTheta > \textbf{0}} & -\log\det\bfTheta+\frac{\rho}{2}\left\Vert\bfTheta+\left(\boldsymbol{Z}_{(k)}-\boldsymbol{Y}_{(k)}+\rho^{-1}\boldsymbol{S}\right)\right\Vert_{F}^{2}.
\end{eqnarray}

%In the second line, we exploit the fact that $\bfTheta$ and $\boldsymbol{S}$ are real and symmetric matrices. We derive the third line by taking into account the linearity of the inner product, omitting the terms that do not affect the minimization with respect to $\bfTheta$. In the 
%fourth line, we make use of the following equality: $\left\Vert\boldsymbol{A}+\boldsymbol{B}\right\Vert^{2}=\left\Vert\boldsymbol{A}\right\Vert^{2}+\left\Vert\boldsymbol{B}\right\Vert^{2}+2\left\langle \boldsymbol{A} \boldsymbol{B}\right\rangle _{F}$.

\section{Proofs of FWER control}

\subsection{Dual problem}

To prove the properties of \eqref{eq:SL1Gslope} we must first consider its dual problem.

\begin{lemma}
Dual problem to the graphical SLOPE \eqref{eq:SL1Gslope} has the following form
\begin{equation} \label{eq:dual_graphical_lasso}
 \hat W=\argmax_{J^D_{\lambda}(W-\bfS) \leq 1} \; \log det (W)\;\;,
\end{equation}
where $J^D_{\lambda}(X)$ is the Gslope dual norm of the symmetric matrix $X$ obtained by applying the regular SLOPE dual norm $J^{\star}_{\lambda}(\cdot)$ to $x^{\star}$ - a vectorized upper triangle of $X$,
$$ J^D_{\lambda}(X) = J^{\star}_{\lambda}(x^{\star})=\max  \left\{ \frac{|x^{\star}|_{(1)}} {\lambda_1}, \dots,  \frac{ \sum_{k=1}^m |x^{\star}|_{(k)}}{ \sum_{k=1}^m \lambda_k} \right\}$$
 (for the derivation of the SLOPE dual norm see e.g. \cite{negrinho2014orbit}) .

\end{lemma}

\begin{proof}

\noindent Let us start by rewriting the Gslope norm in terms of its dual norm. Using the standard formula 
\begin{equation*}
J_{\lambda}(X) = \max_{J^D_{\lambda}(U) \leq 1} \tr{UX}\;\;,
\end{equation*}
\noindent where $U$ is the symmetric matrix,
we obtain the following form on the Gslope optimization problem

\begin{equation*}
\hat X=\argmax_{X \succ 0} \; \log det (X) - \tr{\bfS X} -  \max_{J^D_{\lambda}(U) \leq 1} \tr{UX}.
\end{equation*}

\noindent Using the fact that a trace is an additive function we get

\begin{equation*}
\hat X= \max_{X  \succ  0} \min_{J^D_{\lambda}(U) \leq 1} \; \log det (X) - \tr{X (U+\bfS)}
\end{equation*}

\noindent For  (\ref{eq:SL1Gslope}) the strong duality holds (because the problem is convex and the Slater's condition is satisfied). Therefore we can obtain a dual solution by exchanging the min and the max 

\begin{equation*}
\hat U= \argmin_{J^D_{\lambda}(U) \leq 1}  \max_{X  \succ  0} \; \log det (X) - \tr{X (U+\bfS)}
\end{equation*}

There is a closed formula for the solution of the inner maximization. 
We simply compute the gradient and set it to zero. This yields

\begin{align*}
\boldsymbol{0} &= d (\log det (X) - \tr{X (U+\bfS)}) \\
&= d \log det (X) - d \tr{X (U+\bfS)} \\
&= X^{-1} - (U+\bfS) .
\end{align*}

Then
$ \tr{X (U+\bfS)} = \tr{XX^{-1}} = p$ and
 finally we get the dual problem in the form:

\begin{equation}\label{eq1}
 \hat U= \argmin_{J^D_{\lambda}(U) \leq 1} \; -\log det (U+\bfS) - p\;\;.
\end{equation}

\noindent For the sake of notation let us define $\bfW:=U+\bfS$ and rewrite (\ref{eq1}) in the form

\begin{equation}\label{eq:dual1}
 \hat W= \argmax_{J^D_{\lambda}(\bfW-\bfS) \leq 1} \; \log det (\bfW)\;\;.
\end{equation}

\end{proof}

\noindent The dual problem \eqref{eq:dual1} has an insightful interpretation. 
We maximize $\log det$ of a matrix W, with a constraint
that $\bfW$ is different from the sample covariance matrix $\mS$ by no more than
than 1 in $J^D_{\lambda}$ norm. 
When we solve the graphical SLOPE \eqref{eq:SL1Gslope}, $\bfW$ is our estimate of the covariance matrix.

\subsection{Proof of Theorem \ref{th:Glasso}}.\label{Ap1}

\begin{proof}
Let us sort variables such that the true covariance matrix takes the block diagonal form and 
let us denote by ${\cal M}$ the set of all block diagonal matrices corresponding to the connectivity components of the true graph:
$${\cal M}=\{M: M=\mbox{block diag}\;(D_1,\ldots,D_k), D_i \in \boldsymbol{S}^{k_i}_{+}\}\;\;,$$
where $k_i$ is the number of nodes in $i^{th}$ connectivity component of the true graph and $\boldsymbol{S}^{k}_{+}$ is the set of $k\times k$ dimensional positive-definite matrices.  
Then it clearly holds 
$$\bfSigma \in {\cal M}\;\;\mbox{and}\;\;\bfTheta\in \cal{M}.$$

Since the inverse of the block diagonal matrix is also bloc-diagonal, to prove Theorem \ref{th:Glasso} it is sufficient to show that the solution to the dual problem (\ref{eq:dual_graphical_lasso}) belongs to $\cal{M}$ with the probability larger or equal to $1-\alpha$.

For this aim let us denote by ${\cal H}$ the set of nodes' pairs $(i,j)$, such that $i$ and $j$ belong to different connectivity components of the true graph
\begin{equation}\label{eq:H0}
{\cal H}=\{(i,j):\;i<j\;\mbox{and the edge}\; (i,j)\; \mbox{connects distinct connectivity components of the true graph}\}\;.
\end{equation}

Observe that when $(i,j)$, $i<j$, belongs to the complement of ${\cal H}$, ${\cal H}^C$, then the nodes $i$ and $j$ belong to the same connectivity component.

\begin{itemize}
    \item {Feasibility}

We will at first show that for $\lambda=\lambda^{Bon}$  the set ${\cal M}$ has a nonempty intersection with the set ${\cal F_L}$ of all feasible solutions to the dual Glasso problem, with the probability larger than $1-\alpha$.

For this aim observe that Glasso is a specific instance of Gslope  with the constant $\lambda$ sequence and that the feasibility set for the Glasso dual problem is defined as
$${\cal F_L}=\{W: ||W-S||_{\infty}\leq \lambda\}\;\;.$$

Thus, ${\cal M}$ has a nonempty intersection with the set ${\cal F_L}$ if and only if $|S_{ij}|<\lambda$ for all pairs of nodes from ${\cal H}$.  Therefore
$$P({\cal M} \cap {\cal F_L} \neq \emptyset)= P(\forall (i,j) \in{\cal H},\;\; |S_{ij}|\leq \lambda)=1-P(\exists (i,j) \in{\cal H},\;\; |S_{ij}|> \lambda)\;\;.$$
Now, observe that under the hypothesis that $i$ and $j$ are not correlated it holds:

\begin{equation*}
 \sqrt{n-2} \cfrac{S_{ij}}{\sqrt{S_{ii}S_{jj}-S_{ij}^2}} \sim t(n-2)
\end{equation*}
where $ t(n-2)$ is Student distribution with $n-2$ degrees of freedom.

Thus, for all pairs $(i,j)$ such that $\Sigma_{i,j}=0$, 
$$P(|S_{ij}|\geq \lambda^{Bon})\leq \frac{\alpha}{m}\;\;$$
and 
$$P(\exists (i,j) \in{\cal H}, |S_{ij}|>\lambda^{Bon})\leq \alpha \frac{m_0}{m}\leq \alpha\;\;,$$
where $m_0$ is the cardinality of ${\cal H}$.

\item{Optimality}

To finalize the proof it is enough to observe that in case when 
\begin{equation}\label{asap}
{\cal M} \cap {\cal F_L} \neq \emptyset\;,
\end{equation}
then the optimal solution must be contained in ${\cal M}$.
For this purpose observe that 
\begin{equation}
 \frac{\partial \log |W|}{\partial W} = W^{-1}
\end{equation}

\noindent 
So, the gradient of the dual objective function is equal to zero at off-blog-diagonal elements of $W$ if and only if $W^{-1} \in {\cal M}$. This, together with (\ref{asap}) implies that the optimal solution $\hat W$ to the dual optimization problem must be contained in ${\cal M}$ . This also implies that the solution to the primal problem $\hat \bfTheta= \hat W^{-1}$ is contained in ${\cal M}$.

\end{itemize}

\end{proof}

\subsection{Proof of Theorem \ref{thm:Gslope_fwer_block}}.\label{Ap2}
\begin{proof}

 Based on the proof of Theorem \ref{th:Glasso} it is sufficient to prove that 
 $$P({\cal M} \cap {\cal F}_{SL} \neq \emptyset)\geq 1- \alpha\;\;,$$
where $F_{SL}$ denotes the set of feasible solutions for the dual Gslope problem.

 \noindent

Let us denote by $W^{\star}$ and $\mS^{\star}$ upper triangles of matrices $W$ and $\mS$ vectorized according to the procedure described in Section 2.1. Now, observe that an event
$$\: \forall_k\in \{1,\ldots,m\} \;\;|W^{\star}-\mS^{\star}|_{(k)}  \leq \lambda_k$$ implies  
$$ J^D_{\lambda}(W-\mS) = \max  \left\{ \frac{|W^{\star}-\mS^{\star}|_{(1)}} {\lambda_1}, \dots,  \frac{ \sum_{k=1}^m |W^{\star}-\mS^{\star}|_{(k)}}{ \sum_{k=1}^m \lambda_k} \right\} \leq 1\;\;.$$
Thus
\begin{equation}\label{eq:dual_probability_inequality}
 \{ J^D_{\lambda}(W-\mS) \leq 1 \} \supseteq
\{ \forall_{k=1,\dots,m} \quad |W^{\star}-\mS^{\star}|_{(k)} \leq \lambda_k \}\;\;.
\end{equation}

Let us now consider all edges $(i,j)$ from ${\cal H}$ and sort them according to the magnitude of the corresponding elements of the sample covariance matrix: $|S_{\cal H}|_{(1)}\geq |S_{\cal H}|_{(2)} \geq |S_{\cal H}|_{(m_0)}$. Now, observe that the following solution $W\in{\cal M}$: $W_{ij}=S_{ij}$ for $(i,j)\in {\cal H}^c$ and $W_{ij}=0$ for $(i,j) \in{\cal H}$, is feasible if only for all $k\in{1,\ldots,m_0}$, $|S_{\cal H}|_{(k)}\leq \lambda_k$. 

Thus 
$$P({\cal M} \cap {\cal F_{SL}} \neq \emptyset)\geq P(\forall k \in \{1,\ldots,m_0\},\;\;  |S_{\cal H}|_{(k)}\leq \lambda_k)=1-P(\exists k\in\{1,\ldots,m_0\},\;\; |S_{\cal H}|_{(k)}> \lambda_k)\;\;.$$

Now, observe that for our Holm selection of the sequence of tuning parameter (\ref{Holm_Seq}) and for
all pairs $(i,j)$ such that $\Sigma_{i,j}=0$,
$$P(|S_{ij}|\geq \lambda^{Holm}_k)=\frac{\alpha}{m-k+1}\;\;.$$

Thus 
$$P(\exists k\in\{1,\ldots,m_0\},\;\; |S_{\cal H}|_{(k)}> \lambda_k)\leq P(V_{Hoch}>0)\leq \alpha\;\;,$$
where $V_{Hoch}$ is the number of false discoveries made by the Hochberg multiple testing procedure for testing the set of hypotheses: $H_{ij}: \Sigma_{ij}=0$ based on the statistics $|S_{ij}|.$

\end{proof}

\end{document}